\documentclass[aip,amsmath,amssymb,reprint]{revtex4-1}   

\usepackage{dcolumn}
\usepackage{bm}
\usepackage[utf8]{inputenc}
\usepackage[T1]{fontenc}
\usepackage{mathptmx}

\usepackage{color}
\definecolor{indigo}{RGB}{0,0,120}
\usepackage[colorlinks=true, linkcolor=indigo, citecolor=blue, urlcolor=indigo]{hyperref}

\def\tr{\;{\rm tr}\;}

\def\fl{\noindent}

\newcommand{\pdr}{\partial}
\newcommand{\grad}{{\bf \nabla}}

\newcommand{\beq}{\begin{equation}}
\newcommand{\eeq}{\end{equation}}
\newcommand{\beqs}{\begin{eqnarray}}
\newcommand{\eeqs}{\end{eqnarray}}

\newcommand{\half}{\frac{1}{2}}
\newcommand{\ov}[1]{\frac{1}{#1}}
\newcommand{\fr}[2]{\frac{#1}{#2}}

\def\al{\alpha}	
		
\def\del{\delta}
\def\eps{\epsilon} 
\def\la{\lambda}	
\def\sig{\sigma}

\def\vf{\varphi}		
\def\tht{\theta}	
\def\om{\omega}	
\def\Om{\Omega}

\def\span{\text{span}}
\newcommand{\mR}{{\mathbb{R}}}

\usepackage{graphicx}

\usepackage{subcaption}
\usepackage{ragged2e}
\DeclareCaptionJustification{justified}{\justifying}
\usepackage{calligra} 
\DeclareMathAlphabet{\mathcalligra}{T1}{calligra}{m}{n}
\DeclareFontShape{T1}{calligra}{m}{n}{<->s*[2.2]callig15}{}
\newcommand{\scripty}[1]{\ensuremath{\mathcalligra{#1}}}

\newcount\colveccount  
\newcommand*\colvec[1]{\global\colveccount#1  \begin{pmatrix} \colvecnext} \def\colvecnext#1{#1 \global\advance\colveccount-1
        \ifnum\colveccount>0 \\ \expandafter\colvecnext
        \else \end{pmatrix} \fi}

\newcommand{\err}{\scripty{r}}

\begin{document}




\title[Classical three rotor problem: periodic solutions, stability and chaos \hfill {\tt arXiv:1811.05807}]{
Classical three rotor problem: periodic solutions, stability and chaos}

\author{Govind S. Krishnaswami}
\homepage{https://www.cmi.ac.in/~govind/}
\email{govind@cmi.ac.in}
\author{Himalaya Senapati}
\email{himalay@cmi.ac.in}
\affiliation{Physics Department, Chennai Mathematical Institute,  SIPCOT IT Park, Siruseri 603103, India}

\date{17 Dec, 2019}

\begin{abstract}

Published in Chaos {\bf 29} (12), 123121 (2019) (Editor's pick) \\

\fl This paper concerns the classical dynamics of three coupled rotors: equal masses moving on a circle subject to attractive cosine inter-particle potentials. It is a simpler variant of the gravitational three-body problem and also arises as the classical limit of a model of coupled Josephson junctions. Unlike in the gravitational problem, there are no singularities (neither collisional nor non-collisional), leading to global existence and uniqueness of solutions. In appropriate units, the non-negative energy $E$ of the relative motion is the only free parameter. We find analogues of the Euler-Lagrange family of periodic solutions: pendulum and isosceles solutions at all energies and choreographies up to moderate energies. The model displays order-chaos-order behavior: it is integrable at zero and infinitely high energies but displays a fairly sharp transition from regular to chaotic behavior as $E$ is increased beyond $E_c \approx 4$ and a more gradual return to regularity. The transition to chaos is manifested in a dramatic rise of the fraction of the area of the Hill region of Poincar\'e surfaces occupied by chaotic sections and also in the spontaneous breaking of discrete symmetries of Poincar\'e sections present at lower energies. Interestingly, the above pendulum solutions alternate between being stable and unstable, with the transition energies cascading geometrically from either side at $E = 4$. The transition to chaos is also reflected in the curvature of the Jacobi-Maupertuis metric that ceases to be  everywhere positive when $E$ exceeds four. Examination of Poincar\'e sections also indicates global chaos in a band of energies $(5.33 \lesssim E \lesssim 5.6)$ slightly above this transition.  
\end{abstract}


\maketitle

\begin{quotation}
We study the classical three rotor problem: three equal point masses moving on a circle subject to attractive cosine inter-particle potentials. It arises as the classical limit of a chain of coupled Josephson junctions. Unlike in the gravitational problem, particles can pass through each other without producing singularities. In  center of mass variables, the relative energy $E$ serves as a control parameter. We discover classes of periodic solutions: choreographies up to moderate energies and pendula and breathers at all energies. The system is integrable at zero and infinite energies but displays a fairly sharp transition to chaos around $E \approx 4$, thus providing an instance of the order-chaos-order transition. We find several manifestations of this transition to chaos: (a) a geometric cascade of stable to unstable transition energies in pendula as $E$ approaches four from either side; (b) a transition in the curvature of the Jacobi-Maupertuis metric from being strictly positive to having both signs as $E$ increases beyond four, implying widespread onset of instabilities; (c) a dramatic rise in the fraction of the area of Poincar\'e surfaces occupied by chaotic trajectories and (d) breakdown of discrete symmetries in Poincar\'e sections present at lower energies. Slightly above this transition, we also find numerical evidence for a band of global chaos where we conjecture ergodic behavior. 
\end{quotation}

\scriptsize

\tableofcontents

\normalsize

\section{Introduction}
\label{s:introduction}

The classical gravitational three-body problem \cite{gutzwiller-rmp, gutzwiller-book} is one of the oldest problems in dynamics and was the place where Poincar\'e discovered chaos. It continues to be a fertile area of research with discovery of new phenomena such as choreographies \cite{chenciner-montgomery} and Arnold diffusion \cite{xia-arnold-diffusion}. In this paper, we study the simpler problem of three rotors, where three particles of equal mass $m$ move on a circle subject to attractive cosine inter-particle potentials of strength $g$. The problem of two rotors reduces to that of a simple pendulum while the three rotor system bears some resemblance to a double pendulum as well as to the planar restricted three-body problem. However, unlike in the gravitational three-body problem, the rotors can pass through each other so that there are no collisional singularities. In fact, the boundedness of the potential also ensures the absence of non-collisional singularities leading to global existence and uniqueness of solutions. Despite these simplifications, the dynamics of three (or more) rotors is rich and displays novel signatures of the transition from regular to chaotic motion as the coupling (or energy) is varied.

The quantum version of the $n$-rotor problem is also of interest as it is used to model a chain of coupled Josephson junctions \cite{sondhi-girvin}. Here, the rotor angles are the phases of the superconducting order parameters associated to the segments between junctions. It is well-known that this model for arrays of coupled Josephson junctions is related to the XY model of classical statistical mechanics\cite{sondhi-girvin,wallin-1994} (see  also Appendix \ref{a:n-rotor-from-xy} where we obtain the quantum $n$-rotor problem from the XY model via a partial continuum limit and a Wick rotation). While in the application to the insulator-to-superconductor transition in arrays of Josephson junctions, one is typically interested in the limit of large $n$, here we focus on the classical dynamics of the $n=3$ case. 

The classical $n$-rotor problem also bears some resemblance to the Frenkel-Kontorova (FK) model \cite{braun-kivshar-FK}. The latter describes a chain of particles subject to nearest neighbor harmonic and onsite cosine potentials. Despite having different potentials and target spaces ($\mR^1$ vs $S^1$), the FK and $n$-rotor problems both admit continuum limits described by the sine-Gordon field \cite{braun-kivshar-FK,sachdev}. Though quite different from our model, certain variants of the three rotor problem have also been studied, e.g., (a) chaos in the dynamics of three masses moving on a line segment with periodic boundary conditions subject to harmonic and 1d-Coulombic inter-particle potentials \cite{kumar-miller}, (b) three free but colliding masses moving on a circle and indications of a lack of ergodicity therein \cite{Rabouw-Ruijgrok}, (c) coupled rotors with periodic driving and damping, in connection with mode-locking phenomena \cite{Thouless-Choi} and (d) an open chain of three coupled rotors with pinning potentials and ends coupled to stochastic heat baths, in connection to ergodicity \cite{Eckmann}.

In \S \ref{s:three-rotor-setup}, we begin by formulating the classical three-rotor problem, show absence of singularities and eliminate the center of mass motion to arrive at dynamics on a $2$ dimensional configuration torus parametrized by the relative angles $\vf_1$ and $\vf_2$. In \S \ref{s:dynamics-on-2torus}, we discuss the dynamics on the $\vf_1$-$\vf_2$ torus, find all static solutions for the relative motion and discuss their stability (see Fig. \ref{f:static-solutions-3rotors}). The system is also shown to be integrable at zero and infinitely high relative energies $E$ (compared to the coupling $g$) due to the emergence of additional conserved quantities. Furthermore, using Morse theory, we discover changes in the topology of the Hill region of the configuration space at $E = 0$, $4g$ and $4.5g$ (see Fig. \ref{f:topology-hill-region}).

 In \S \ref{s:reduction-one-dof}, we use consistent reductions of the equations of motion to one degree of freedom to find two families of periodic solutions at all energies (pendula and isosceles breathers, see Fig. \ref{f:periodic-soln}). This is analogous to how the Euler and Lagrange solutions of the 3 body problem arise from suitable Keplerian orbits. We investigate the stability of the pendula and breathers by computing their monodromies. Notably, we find that the stability index of pendula becomes periodic on a log scale as $E \to 4g^\pm$ and shows an accumulation of stable to unstable transition energies at $E = 4g$ (see Fig. \ref{f:monodromy-evals}). In other words, the largest Lyapunov exponent switches from positive to zero infinitely often with the widths of the (un)stable windows asymptotically approaching a geometric sequence as the pendulum energy approaches $4g$. This accumulation bears an interesting resemblance to the Efimov effect \cite{efimov-original} as discussed in \S \ref{s:discussion} and to the cascade of period doubling bifurcations in unimodal maps \cite{logistic-map}.

In \S \ref{s:JM-approach}, we reformulate the dynamics on the $\vf_1$-$\vf_2$ torus as geodesic flow with respect to the Jacobi-Maupertuis metric. We prove in Appendix \ref{a:positivity-of-JM-curvature} that the scalar curvature is strictly positive on the Hill region for $0 \leq E \leq 4g$ but acquires both signs above $E = 4g$ (see Fig. \ref{f:curvature-3rotors-phi1-phi2-torus}) indicating widespread geodesic instabilities as $E$ crosses $4g$. In \S \ref{s:poincare-section}, we examine Poincar\'e sections and observe a marked transition to chaos in the neighborhood of $E = 4g$ as manifested in a rapid rise of the fraction of the area of the energetically allowed `Hill' region occupied by chaotic sections (see Fig. \ref{f:chaos-vs-egy}). This is accompanied by a spontaneous breaking of two discrete symmetries present in Poincar\'e sections below this energy (see Figs. \ref{f:psec-egy=2-3} and \ref{f:psec-egy-near-4}). This transition also coincides with the accumulation of stable to unstable transition energies of the pendulum family of periodic solutions at $E = 4g$. Slightly above this energy, we find a band of global chaos $5.33g \lesssim E \lesssim 5.6g$, where the chaotic sections fill up the entire Hill region on all Poincar\'e surfaces, suggesting ergodic behavior (see Fig. \ref{f:global-chaos}). In \S \ref{s:choreographies}, we derive a system of delay differential and algebraic equations for periodic choreography solutions of the three rotor problem. We discover three families of choreographies. The first pair are uniformly rotating versions of two of the static solutions for the relative motion. The third family is non-rotating, stable and exists for all relative energies up to the onset of global chaos (see Fig. \ref{f:choreo-time-vs-egy-phi1-phi2-plot}). It is found by a careful examination of Poincar\'e sections. Finally, we prove that choreographies cannot exist for arbitrarily high relative energies. We conclude with a discussion in \S \ref{s:discussion}. Appendix \ref{a:estimate-f} summarizes the numerical method employed to estimate the fraction of chaos on Poincar\'e surfaces. A preliminary version of this paper was presented at the Conference on Nonlinear Systems and Dynamics, New Delhi, October 2018 \cite{gskhs-cnsd-3rotor}.

\section{Three coupled classical rotors}
\label{s:three-rotor-setup}

We study a periodic chain of three identical rotors of mass $m$ interacting via cosine potentials. The Lagrangian is 
	\beq
	L = \sum_{i=1}^3  \left\{ \half { m r^2} \dot\tht_i^2 - g [1 - \cos\left(\tht_i-\tht_{i+1} \right) ] \right\}
	\eeq
with $\tht_4 \equiv \tht_1$. Here, $\tht_i$ are $2\pi$-periodic coordinates on a circle of radius $r$. Though we only have nearest neighbor  interactions, each pair interacts as there are only three rotors. We consider the `ferromagnetic' case where the coupling $g > 0$ so that the rotors attract each other. Unlike in the gravitational three-body problem, the inter-rotor forces vanish when a pair of them coincide so that rotors can `pass' through each other: this is physically reasonable since they occupy distinct sites. The equations of motion for $i = 1$, $2$ and $3$ (with $\tht_0 \equiv \tht_3$ and $\tht_1 \equiv \tht_4$) are
	\beq
	m r^2 \ddot \tht_i = g \sin (\tht_{i-1}-\tht_i) - g \sin (\tht_i-\tht_{i+1}).
	\eeq
This is a system with three degrees of freedom, the configuration space is a 3-torus $0 \leq \tht_i \leq 2\pi$. The conjugate angular momenta are $\pi_i = m r^2 \dot \tht_i$ and the Hamiltonian is
	\beq
	H = \sum_{i =1}^3  \left\{ \fr{\pi_i^2}{2 m r^2} + g [1- \cos\left(\tht_i-\tht_{i+1} \right) ] \right\}.
	\eeq
Hamilton's equations 
	\beq
	\dot \tht_i = \fr{\pi_i}{mr^2} \;\; \text{and} \;\; \dot \pi_i = g [\sin (\tht_{i-1}-\tht_i) - \sin (\tht_i-\tht_{i+1})]
	\label{e:hamilton-eom-theta}
	\eeq
define a smooth Hamiltonian vector field on the 6d phase space of the three-rotor problem. The additive constant in $H$ is chosen so that the minimal value of energy is zero. This system has three independent dimensionful physical parameters $m$, $r$ and $g$ that can be scaled to one by a choice of units. However, once such a choice of units has been made, all other physical quantities (such as $\hbar$) have definite numerical values. This circumstance is similar to that in the Toda model \cite{gutzwiller-book}. As discussed in Appendix \ref{a:n-rotor-from-xy}, the quantum $n$-rotor problem, which models a chain of Josephson junctions, also arises by Wick-rotating a partial continuum limit of the XY model on a lattice with nearest neighbor ferromagnetic coupling $J$, $n$ horizontal sites and horizontal and vertical spacings $a$ and $b$ (\ref{e:classical-action-intermediate-problem}). The above parameters are related to those of the Wick-rotated XY model via $m = J/c^2$, $r = \sqrt{L b^2/a}$ and $g = J L/a$ where $L = n a$ and $c$ is a speed associated to the Wick rotation to time.

The Hamiltonian vector field (\ref{e:hamilton-eom-theta}) is non-singular everywhere on the phase space. In particular, particles may pass through one another without encountering collisional singularities. Though the phase space is not compact, the constant energy $(H = E)$ hypersurfaces are compact 5d submanifolds without boundaries. Indeed, $0 \leq \tht_i \leq 2\pi$ are periodic coordinates on the compact configuration space $T^3$. Moreover, the potential energy is non-negative so that $\pi_i^2 \leq 2 m r^2 E$. Thus, the angular momenta too have finite ranges. Consequently, we cannot have `non-collisional singularities' where the (angular) momentum or position diverges in finite time. Solutions to the initial value problem (IVP) are therefore expected to exist and be unique for all time. 

Alternatively, the Hamiltonian vector field is globally Lipschitz since it is everywhere differentiable and its differential bounded in magnitude on account of energy conservation. This means that there is a common Lipschitz constant on the energy hypersurface, so that a unique solution to the IVP is guaranteed to exist for $0 \leq t \leq t_0$ where $t_0 > 0$ is independent of initial condition (IC). Repeating this argument, the solution can be extended for $t_0 \leq t \leq 2t_0$ and thus can be prolonged indefinitely in time for any IC, implying global existence and uniqueness \cite{global-existence-ode}. 

In \S \ref{s:JM-approach}, we will reformulate the dynamics as geodesic flow on $T^2$ (or $T^3$ upon including center of mass motion, see below), which must be geodesically complete as a consequence. For $E > 4.5g$, this is expected on account of compactness and lack of boundary of the energetically allowed Hill region. For $E < 4.5 g$, though the trajectories can (in finite time) reach the Hill boundary, they simply turn around. Examples of such trajectories are provided by the $\vf_1 = 0$ pendulum solutions described in \S \ref{s:pendulum-soln}.

\subsection{Center of Mass (CM) and relative coordinates}

 It is convenient to define the CM and relative angles
	\beq
	\vf_0 = \fr{\tht_1 + \tht_2 + \tht_3}{3}, 
	\;\; \vf_1 = \tht_1 - \tht_2 \;\; \text{and} \;\;  \vf_2 = \tht_2 - \tht_3
	\eeq
or equivalently,
	\beqs
	\tht_1 &=& \vf_0 + \fr{2\vf_1}{3} + \fr{\vf_2}{3}, \quad \tht_2 = \vf_0 - \fr{\vf_1}{3} + \fr{\vf_2}{3} \quad \text{and} \cr
	 \tht_3 &=& \vf_0 - \fr{\vf_1}{3} - \fr{2\vf_2}{3}.
	\eeqs
As a consequence of the $2\pi$-periodicity of the $\tht$s, $\vf_0$ is $2\pi$-periodic while $\vf_{1,2}$ are $6\pi$-periodic. However, the cuboid ($0 \leq \vf_0 \leq 2\pi$, $0 \leq \vf_{1,2} \leq 6\pi$) is a nine-fold cover of the fundamental cuboid $0 \leq \tht_{1,2,3} \leq 2\pi$. In fact, since the configurations $(\vf_0, \vf_1 - 2\pi, \vf_2)$, $(\vf_0, \vf_1, \vf_2 + 2\pi)$ and $(\vf_0 +2\pi/3, \vf_1, \vf_2)$ are physically identical, we may restrict $\vf_{1,2}$ to lie in $[0,2\pi]$. Here, the $\vf_i$ are not quite periodic coordinates on $T^3 \equiv [0,2\pi]^3$. Rather, when $\vf_1 \mapsto \vf_1 \pm 2\pi$ or $\vf_2 \mapsto \vf_2 \mp 2\pi$, the CM variable $\vf_0 \mapsto \vf_0 \pm 2\pi/3$. In these coordinates, the  Lagrangian becomes $L = T - V$ where
	\beqs
	T &=&  \fr{3}{2} m r^2 \dot \vf_0^2 + \fr{1}{3} m r^2 \left[  \dot \vf_1^2 + \dot \vf_2^2 + \dot \vf_1\dot \vf_2 \right] \;\; \text{and} \cr
 V &=& g \left[ 3 - \cos \vf_1  - \cos \vf_2  - \cos( \vf_1 +  \vf_2) \right],  
 	\label{e:lagrangian-phi-coords-3rotors}
 	\eeqs
with the equations of motion (EOM) $3 m r^2 \ddot \vf_0 = 0$, 
	\beq
	m r^2 \left(2 \ddot \vf_1 + \ddot \vf_2 \right) = - 3 g \left[ \sin \vf_1 + \sin ( \vf_1 + \vf_2) \right] \;\; \text{and} \;\; 1 \leftrightarrow 2. 
	\label{e:3rotors-cm-rel-coords-lagrangian-eom-2nd-order}
	\eeq
The evolution equations for $\vf_1$ (and $\vf_2$ with $1 \leftrightarrow 2$) may be rewritten as
	\beq
	 m r^2 \ddot \vf_1 =  - g \left[ 2 \sin \vf_1 - \sin \vf_2 + \sin(\vf_1 + \vf_2)\right].	
	\label{e:3rotors-EOM-ph1ph2}
	\eeq
Notice that when written this way, the `force' on the RHS isn't the gradient of any potential, as the equality of mixed partials would be violated. The (angular) momenta conjugate to $\vf_{0,1,2}$ are $p_0 = 3 m r^2 \dot \vf_0$,
	\beq
	p_1 = \fr{m r^2}{3} (2 \dot \vf_1 + \dot \vf_2) \;\;\; \text{and}
	\;\;\; p_2 = \fr{m r^2}{3} ( \dot \vf_1 + 2 \dot \vf_2).
	\eeq
The remaining three EOM on phase space are $\dot p_0 = 0$ (conserved due to rotation invariance),
	\beqs
	\dot p_1 &=& - g \left[ \sin \vf_1 + \sin (\vf_1 + \vf_2) \right] \;\; \text{and} \cr
	\dot p_2 &=& - g \left[ \sin \vf_2 + \sin (\vf_1 + \vf_2) \right].
	\label{e:3rotors-EOM-p0p1p2}
	\eeqs
The EOM admit a conserved energy which is a sum of CM, relative kinetic and potential energies:
	\beq
	E =   \fr{3}{2} m r^2 \dot \vf_0^2 + \fr{1}{3} m r^2 \left[  \dot \vf_1^2 + \dot \vf_2^2 + \dot \vf_1\dot \vf_2 \right] + V(\vf_1, \vf_2).
 	\label{e:egy-3rotors-phi1-phi2-coords}
	\eeq
The above EOM are Hamilton's equations $\dot f = \{f, H\}$ for canonical Poisson brackets (PBs) $\{ \vf_i, p_j \} = \del_{ij}$ with the Hamiltonian
	\beq
	H = \fr{p_0^2}{6m r^2} + \fr{p_1^2 + p_2^2 - p_1 p_2}{m r^2} + V(\vf_1, \vf_2).
	\eeq

\subsection{Analogue of Jacobi coordinates}
\label{s:jacobi-coordinates}

 Jacobi coordinates for the three rotor problem are $\vf_0$ and $\vf_\pm = (\vf_1 \pm \vf_2)/2$. Unlike in the CM and relative coordinates, the kinetic energy as a quadratic form in velocities is diagonal. Indeed, $L = T - V$ where
	\beqs
	T &=& \fr{3}{2} m r^2 \dot \vf_0^2 + m r^2 \dot \vf_+^2  + \fr{1}{3} m r^2 \dot \vf_-^2 \quad \text{and} \cr 
	V &=& g \left(3 - 2 \cos \vf_- \cos \vf_+ - \cos 2\vf_+  \right).
	\eeqs
The conjugate momenta $p_0$ and $p_\pm =  p_1 \pm p_2$ are proportional to the velocities and the EOM are
	\beqs
	\dot p_0 = 0, \quad \dot p_+ &=& -2g \sin \vf_+ \left( \cos \vf_- + 2 \cos \vf_+ \right) \cr
	 \text{and} \quad
	\dot p_- &=& - 2g \cos \vf_+ \sin \vf_-.
	\eeqs
The fundamental domain which was the cube $0 \leq \vf_{0,1,2} \leq 2\pi$ now becomes the cuboid ($0 \leq \vf_0 \leq 2\pi$, $0 \leq \vf_+ \leq 2\pi$, $0 \leq \vf_- \leq \pi$). As before, though $\vf_\pm$ are periodic coordinates on a 2-torus, $\vf_{0,\pm}$ are not quite periodic coordinates on $T^3$. The transformation of the CM variable $\vf_0$ under $2\pi$-shifts of $\vf_{1,2}$ discussed above may be reformulated as follows. When crossing the segments $\vf_+ + \vf_- = 2\pi$ from left to right or $\vf_+ - \vf_- = 0$ from right to left, $\vf_0$ increases by $2\pi/3$ [and $\vf_0 \mapsto \vf_0 - 2\pi/3$ when the segments are crossed in the opposite direction].

\section{Dynamics on the $\vf_1$-$\vf_2$ torus}
\label{s:dynamics-on-2torus}

The dynamics of $\vf_{1}$ and $\vf_2$  (or equivalently that of $\vf_\pm$) decouples from that of the CM coordinate $\vf_0$. The former may be regarded as periodic coordinates on the 2-torus $[0,2\pi] \times [0,2\pi]$. On the other hand, $\vf_0$, which may be regarded as a fibre coordinate over the $\vf_{1,2}$ base torus, evolves according to 
	\beq
	\vf_0 = \fr{p_0 t}{3mr^2} + \vf_0(0) + \fr{2 \pi}{3} (n_2 - n_1) \quad \text{mod}\; 2\pi.
	\eeq
Here, $n_{1,2}$ are the `greatest integer winding numbers' of the trajectory around the cycles of the base torus. If a trajectory goes continuously from $\vf^i_{1,2}$ to $\vf^f_{1,2}$ (regarded as real rather than modulo $2\pi$), then the greatest integer winding numbers are defined as $n_{1,2} = [(\vf^f_{1,2}-\vf^i_{1,2})/2\pi]$. 

Consequently, we may restrict attention to the dynamics of $\vf_1$ and $\vf_2$. The equations of motion on the corresponding 4d phase space (the cotangent bundle of the 2-torus) are
	\beq
	\dot \vf_{1} = (2 p_{1} - p_{2})/m r^2, \;\;\;
	\dot p_1 = - g \left[ \sin \vf_1 + \sin ( \vf_1 + \vf_2) \right]
	\label{e:3rotors-CM-EOM-phasespace}
	\eeq
and $1\leftrightarrow 2$. These equations define a singularity-free vector field on the phase space. They follow from the canonical PBs with Hamiltonian given by the relative energy
	\beq
	H_\text{rel} = \fr{p_1^2 + p_2^2 - p_1 p_2}{m r^2} + V(\vf_1, \vf_2).
	\label{e:hamiltonian-3rotor-phi1phi2space-full}
	\eeq
These equations and Hamiltonian are reminiscent of those of the planar double pendulum with the Hamiltonian
	\beq
	H_{\rm dp} = \fr{p_1^2 - 2 c_{12}\, p_1\, p_2 + 2 p_2^2}{2 m l^2 (2 -c_{12}^2)} - m g l ( 2 \cos \tht_1 + \cos \tht_2)
	\eeq
where $\tht_{1,2}$ are the angles between the upper and the lower rods (each of length $l$) and the vertical and $c_{12} = \cos (\tht_1-\tht_2)$.

\subsection{Static solutions and their stability}
\label{s:static-sol-and-stability}

\begin{figure*}
	\centering
	\begin{subfigure}[t]{3cm}
		\centering
		\includegraphics[width=3cm]{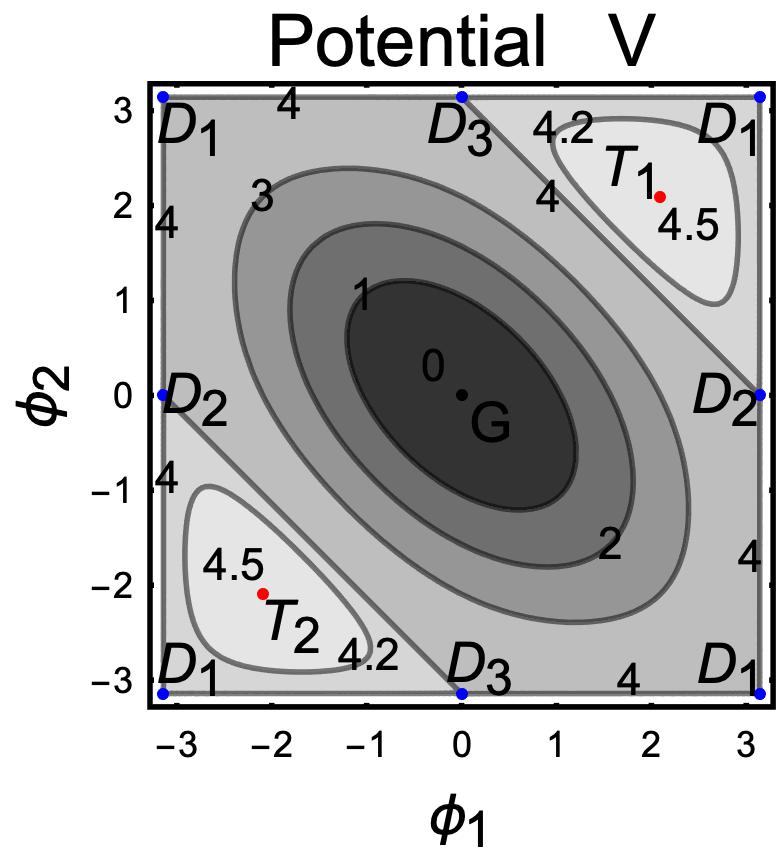}
		\caption{Contours of $V$.}
		\label{f:potential-contour-plot}
	\end{subfigure}
	\quad
	\begin{subfigure}[t]{3cm}
		\centering
		\includegraphics[width=2cm]{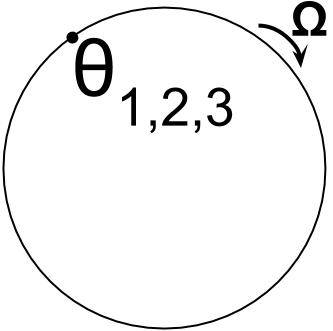}
		\caption{  Ground state G.}
		\label{f:3rotors-ground-state}	
	\end{subfigure}	
	\quad
	\begin{subfigure}[t]{3.2cm}
		\centering
		\includegraphics[width=2cm]{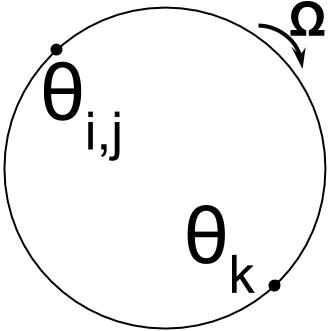}
		\caption{  Diagonal states D.}
		\label{f:3rotors-first-excited}	
	\end{subfigure}	
	\quad
	\begin{subfigure}[t]{3cm}
		\centering
		\includegraphics[width=2cm]{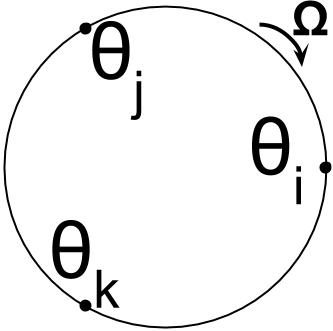}
		\caption{  Triangle states T.}
		\label{f:3rotors-second-excited}
	\end{subfigure}	
	\caption{\small (a) Potential energy $V$ in units of $g$ on the $\vf_1$-$\vf_2$ configuration torus with its extrema (locations of static solutions G, D and T) indicated. The contours also encode changes in topology of the Hill region ($V \leq E$) when $E$ crosses $E_{\rm G} = 0$, $E_{\rm D} = 4g$ and $E_{\rm T} = 4.5g$. (b, c, d) Uniformly rotating three-rotor solutions obtained from G, D and T. Here, $i,j$ and $k$ denote any permutation of the numerals $1$, $2$ and $3$. (b) and (d) are the simplest examples of choreographies discussed in \S \ref{s:choreographies}.}
	\label{f:static-solutions-3rotors}
\end{figure*}

Static solutions for the relative motion correspond to zeros of the vector field where the force components in (\ref{e:3rotors-CM-EOM-phasespace}) vanish: $p_1 = p_2 = 0$ and
	\beq
	\sin \vf_1 + \sin (\vf_1+\vf_2) = \sin \vf_2 + \sin (\vf_1 + \vf_2) = 0.
	\eeq 
In particular, we must have $\vf_1 = \vf_2$ or $\vf_1 = \pi - \vf_2$. When $\vf_1 = \vf_2$, the force components are both equal to $\sin \vf_1 (1 + 2\cos \vf_1)$ which vanishes at the following configurations:
	\beq
	(\vf_1, \vf_2)\; = \;(0,0), \; \left(\pi, \pi\right) \; \text{and} \; \left(\pm {2\pi}/{3}, \pm {2\pi}/{3}\right).
	\eeq
On the other hand, if $\vf_1 = \pi - \vf_2$, we must have $\sin \vf_1 = 0$  leading to two more static configurations $(0, \pi)$ and $(\pi, 0)$. Thus we have six static solutions which we list in increasing order of (relative) energy:
	\beqs
& E=0: G(0,0), \quad E=4g: D_1(\pi, \pi), D_2(\pi, 0), D_3(0, \pi) \cr
& \text{and} \quad E = 9g/2: T_{1,2}(\pm {2\pi}/{3},\pm{2\pi}/{3}). 
	\eeqs
Below, we clarify their physical meaning by viewing them as uniformly rotating three body configurations.

\subsubsection{Uniformly rotating 3-rotor solutions from G, D and T} 
\label{s:rotating-static-sol}

If we include the uniform rotation of the CM angle ($\dot \vf_0 = \Om$ is arbitrary), these six solutions correspond to the following uniformly rotating rigid configurations of 3-rotors (see Fig.~\ref{f:static-solutions-3rotors}):  (a) the ferromagnetic ground state G where the three particles coalesce ($\tht_1 = \tht_2 = \tht_3$), (b) the three `diagonal' `anti ferromagnetic N\'eel'  states D where two particles coincide and the third is diametrically opposite ($\tht_1 = \tht_2 = \tht_3 + \pi$ and cyclic permutations thereof) and (c) the two `triangle' `spin wave' states T where the three bodies are equally separated ($\tht_1 = \tht_2 + 2\pi/3 = \tht_3 + 4\pi/3$ and $\tht_2 \leftrightarrow \tht_3$). 

\subsubsection{Stability of static solutions}

 The linearization of the EOM (\ref{e:3rotors-EOM-ph1ph2}) for perturbations to G, D and T ($\vf_{1,2} = \bar \vf_{1,2} + \delta\vf_{1,2}(t)$) take the form
	\beqs
	 &mr^2 \fr{d^2}{dt^2} \colvec{2}{\delta\vf_1} {\delta\vf_2} = -g A \colvec{2}{\delta\vf_1} {\delta\vf_2}
	\quad \text{where} \quad
	A_{\text{G}} = 3 I, \cr 
	&A_{\text{D$_3$}(0,\pi)}=\colvec{2}{1 & 0}{-2 & -3}, \quad
	 A_{\text{D$_2$}(\pi,0)} = \colvec{2}{-3 & -2}{0 & 1}, \cr
	& A_{\text{D$_1$}(\pi, \pi)}=\colvec{2}{-1 & 2}{2 & -1}\quad \text{and}
	\quad
	A_{\rm T}= -3I/2.
	\label{e:EOM-ph1ph2-perturbation-static-solution}
	\eeqs
Here $I$ is the $2 \times 2$ identity matrix. Perturbations to G are stable and lead to small oscillations with equal frequencies $\om_0 = \sqrt{3g/mr^2}$. The saddles D have one stable direction with frequency $\om_0/\sqrt{3}$ and one unstable eigendirection with growth rate $\om_0$. On the other hand, both eigendirections around T are unstable with growth rate $\om_0/\sqrt{2}$.

\subsection{Changes in topology of Hill region with growing energy}
\label{s:topology-hill-region}

\begin{figure*}
	\centering
	\begin{subfigure}[c]{9.5cm}
		\centering
		\includegraphics[width=10cm]{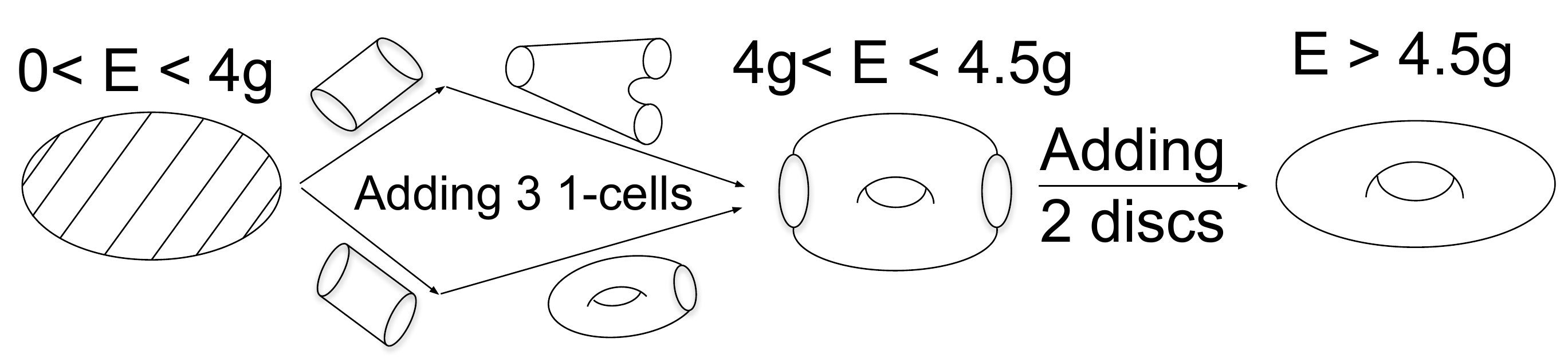}
		\caption{ }
		\label{f:topology-hill-region-a}
	\end{subfigure}	
	\qquad \quad
	\begin{subfigure}[c]{5cm}
		\centering
		\includegraphics[width=6cm]{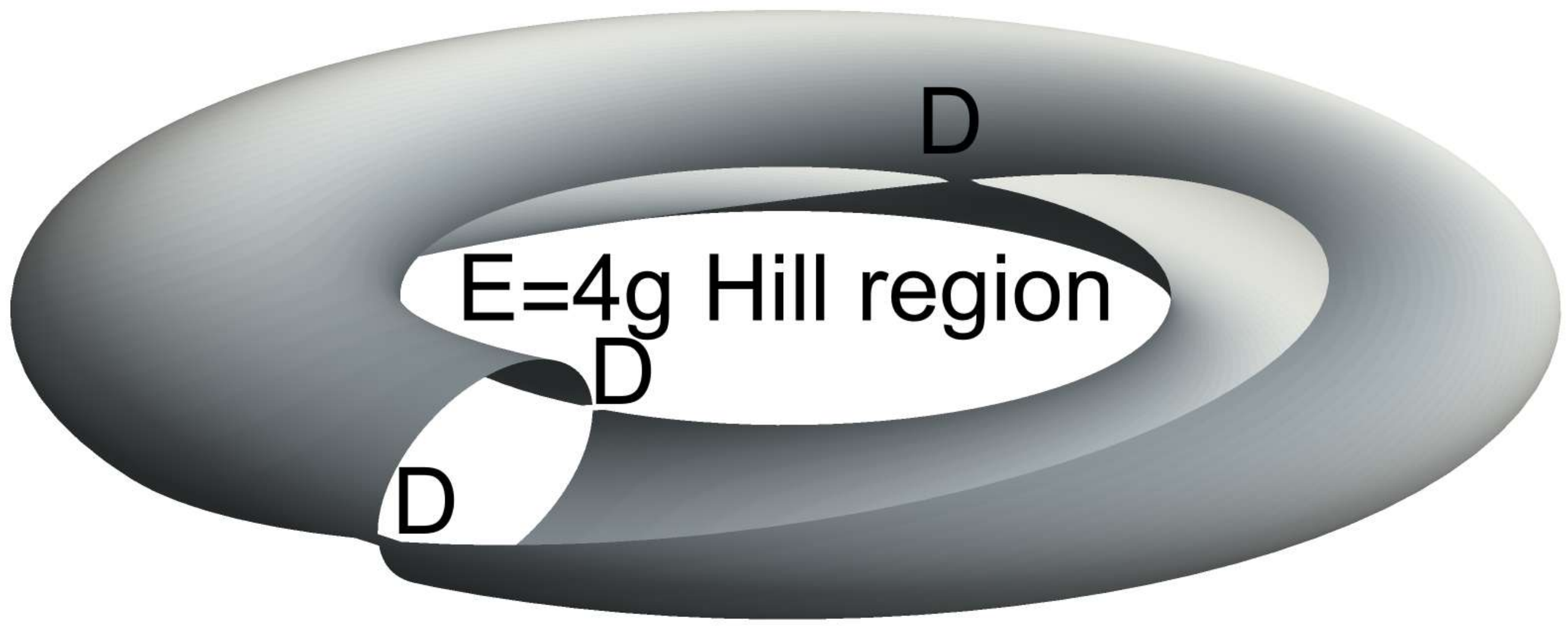}
		\caption{ }
		\label{f:topology-hill-region-b}
	\end{subfigure}	
	\caption{ (a) Topology of Hill region of configuration space $(V(\vf_1, \vf_2) \leq E)$ showing transitions at $E = 4g$ and $4.5g$ as implied by Morse theory (see \S \ref{s:dynamics-on-2torus}). (b) The Hill region for $E = 4g$ is not quite a manifold; its boundary consists of 3 non-contractible closed curves on the torus meeting at the D configurations.} 
	\label{f:topology-hill-region}
\end{figure*}

The Hill region of possible motions ${\cal H}_E$ at energy $E$ is the subset $V(\vf_1, \vf_2) \leq E$ of the $\vf_1$-$\vf_2$ configuration torus. The topology of the Hill region for various energies can be read-off from Fig.~\ref{f:potential-contour-plot}. For instance, for $0 < E < 4g$, ${\cal H}_E$ is a disc while it is the whole torus for $E > 4.5 g$. For $4g < E < 4.5g$, it has the topology of a torus with a pair of discs (around T$_1$ and T$_2$) excised (see also Fig. \ref{f:curvature-3rotors-phi1-phi2-torus}). These changes in topology are confirmed by Morse theory \cite{milnor} if we treat $V$ as a real-valued Morse function, since its critical points are non-degenerate (non-singular Hessian). In fact, the critical points of $V$ are located at G (minimum with index 0), D$_{1,2,3}$ (saddles with indices 1) and T$_{1,2}$ (maxima with indices 2). Thus, the topology of ${\cal H}_E$ can change only at the critical values $E_G = 0, E_D = 4g$ and $E_T = 4.5g$ (see Fig. \ref{f:topology-hill-region-a}). The topological transition from ${\cal H}_{E < 4g}$ (disc) to ${\cal H}_{4g < E < 4.5g}$ (torus with two discs excised) can be achieved by the successive addition of three 1-cells to the disc (proceeding either via a cylinder and a pair of pants or a cylinder and a torus with one disc excised). Similarly, one arrives at the toroidal Hill region for $E > 4.5 g$ by sewing two 2-cells to cover the excised discs of ${\cal H}_{4g < E < 4.5g}$ as depicted in  Fig. \ref{f:topology-hill-region-a}. At the critical value $E = 0$, ${\cal H}_E$ shrinks to a point while at $E = 4.5 g$, it is a twice-punctured torus. Fig.~\ref{f:topology-hill-region-b} illustrates the Hill region at the critical value $E = 4g$ where alone it is not quite a manifold.


\subsection{Low and high energy limits}
\label{s:low-and-high-egy-limit}

In the CM frame, the three-rotor problem (\ref{e:3rotors-CM-EOM-phasespace}) has a 4-dimensional phase space but possesses only one known conserved quantity (\ref{e:hamiltonian-3rotor-phi1phi2space-full}). However, an extra conserved quantity emerges at zero and infinitely high energies:

(a) For $E \gg g$, the  kinetic energy dominates and $H \approx (p_1^2 - p_1 p_2 + p_2^2)/mr^2$. Here $\vf_{1,2}$ become cyclic coordinates and  $p_{1,2}$ are both approximately conserved.

(b) For $E \ll g$, the system executes small oscillations around the ground state G $(\vf_{1,2} \equiv 0)$. The quadratic approximation to the Lagrangian (\ref{e:lagrangian-phi-coords-3rotors}) for relative motion is
	\beq
	L_{\rm low} = \fr{mr^2}{3} \left[\dot \vf_1^2 + \dot \vf_2^2 + \dot \vf_1 \dot \vf_2 \right] - g \left(\vf_1^2 + \vf_2^2 + 
\vf_1 \vf_2 \right).
	\label{e:Lagrangian-low-egy}
	\eeq
The linear equations of motion for $\vf_1$ and $\vf_2$ decouple,
	\beq
	m r^2 \ddot \vf_{1} = - 3 g \vf_{1} \quad \text{and} \quad m r^2 \ddot \vf_{2} = - 3 g \vf_{2}
	\label{e:decoupled-sho-eqn-3rotors-low-egy-limit}
	\eeq
leading to the separately conserved normal mode energies $E_{1,2} = \left( m r^2 \dot \vf_{1,2}^2 + 3 g \vf_{1,2}^2 \right)/2$. The equality of frequencies implies that any pair of independent linear combinations of $\vf_1$ and $\vf_2$ are also normal modes. Of particular significance are the Jacobi-like variables $\vf_\pm = (\vf_1 \pm \vf_2)/2$ that diagonalize the kinetic and potential energy quadratic forms:
	\beq
	L_{\rm low} = m r^2 \dot \vf_+^2 - 3g \vf_+^2 + m r^2 \dot \vf_-^2/3 - g \vf_-^2.
	\eeq
Though (\ref{e:decoupled-sho-eqn-3rotors-low-egy-limit}) are simply the EOM for a pair of decoupled oscillators, the Lagrangian and Poisson brackets $\{ \cdot , \cdot \}$ inherited from the non-linear theory are different from the standard ones. With conjugate momenta $p_{1,2} = (mr^2/3)  (2 \dot \vf_{1,2} + \dot \vf_{2,1})$, the Hamiltonian corresponding to (\ref{e:Lagrangian-low-egy}) is
	\beq
	H_{\rm low} = \frac{p_1^2 - p_1 p_2 + p_2^2}{ mr^2} +  g \left(\vf_1^2 + \vf_2^2 + \vf_1 \vf_2 \right).
	\label{e:low-egy-hamiltonian-phi1-phi2-3rotors}
	\eeq
Note that $p_{1,2}$ differ from the standard momenta $p_{1,2}^{\rm s} = m r^2 \dot \vf_{1,2}$ whose PBs are now non-canonical,  $\{\vf_i, p_j^{\rm s}\} = -1 + 3 \delta_{ij}$.

\subsubsection{Three low-energy constants of motion}

$H_{\rm low}$ and the normal mode energies
	\beq
	H_{1,2} = \left( 2 p_{1,2} - p_{2,1} \right)^2/2 m r^2 + {3g}\vf_{1,2}^2/2
	\label{e:low-egy-normal-modes-3rotors-phi1-phi2}
	\eeq
are three independent constants of motion in the sense that the corresponding 1-forms $dH$, $dH_1$ and $dH_2$ are generically linearly independent ($dH \wedge dH_1 \wedge dH_2 \not \equiv 0$ on the 4d phase space). On the other hand, we also have a conserved `angular momentum' $L_z = m r^2 (\vf_1 \dot \vf_2 - \vf_2 \dot \vf_1)$ corresponding to the rotation invariance of the decoupled oscillators in (\ref{e:decoupled-sho-eqn-3rotors-low-egy-limit}). It turns out that $H_{\rm low}$ may be expressed as
	\beq
	H_{\rm low} = \fr{2}{3}\left[ H_1 + H_2 +\sqrt{ H_1 H_2 - (3 g/4 m r^2) L_z^2} \right].
	\eeq
The low energy phase trajectories lie on the common level curves of $H_{\rm low}$, $H_1$ and $H_2$. Though $H_1$ and $H_2$ are conserved energies of the normal modes, they do not Poisson commute. In fact, the Poisson algebra of conserved quantities is $\{H_{1,2}, H_{\rm low}\} = \{L_z, H_{\rm low} \} = 0$,
	\beqs
	\{H_1, H_2\} &=& -3 g L_z/m r^2 \quad \text{and} \cr
	 \{L_z, H_{1,2}\} &=& \pm 2(3H_{\rm low} - 2H_{1,2} - H_{2,1}).
	\eeqs
It is also noteworthy that the integrals $H_1 + H_2$ and $H_1 H_2 - 3g L_z^2/(4mr^2)$ are in involution.

\section{Reduction to one degree of freedom} 
\label{s:reduction-one-dof}

\begin{figure}	
	\centering
	\begin{subfigure}[t]{2.5cm}
		\centering
		\includegraphics[width=2.5cm]{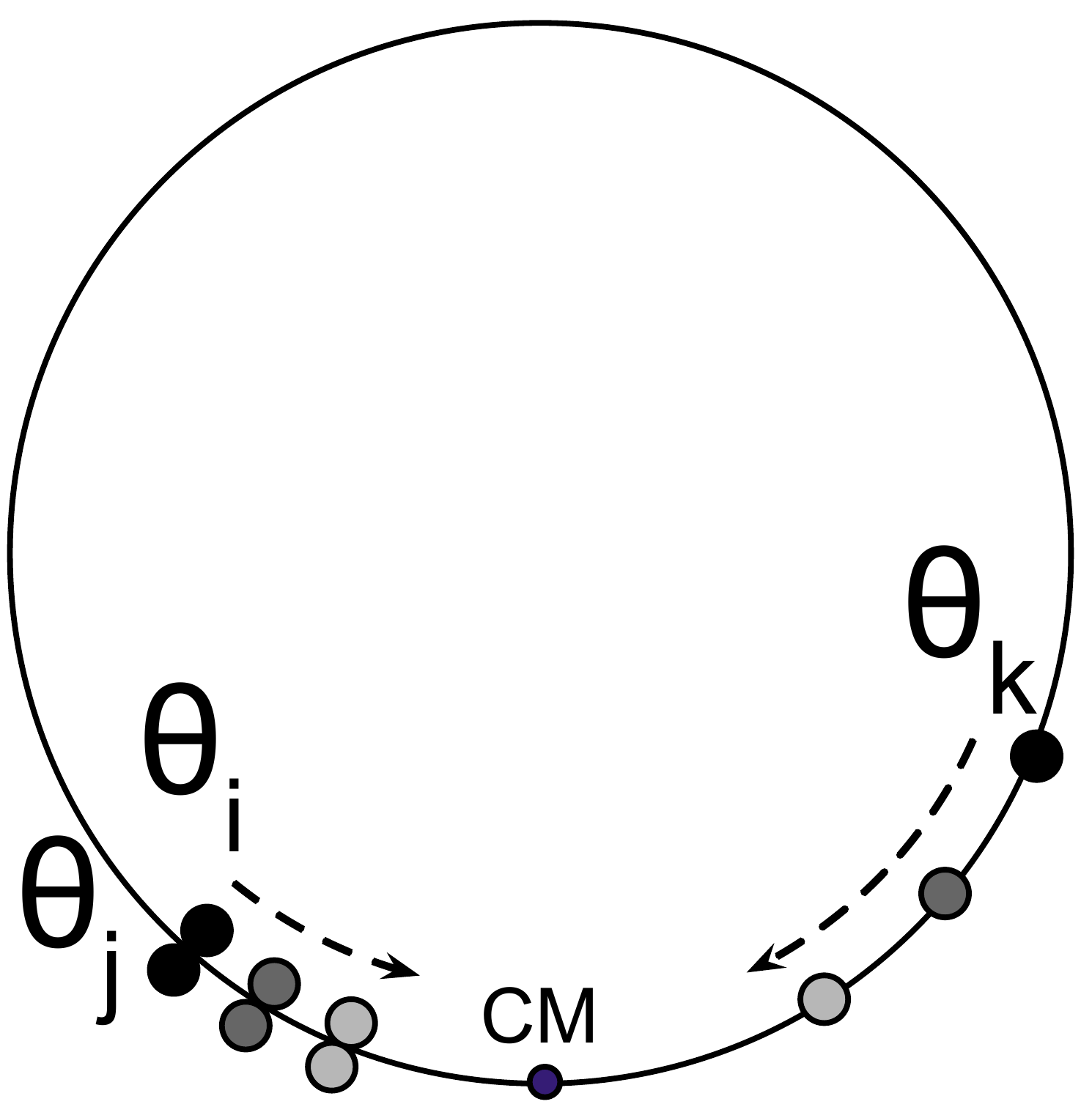}
	\subcaption{  Pendula}
	\label{f:pendula}
	\end{subfigure}
\qquad
	\begin{subfigure}[t]{3.2cm}
		\centering
		\includegraphics[width=2.5cm]{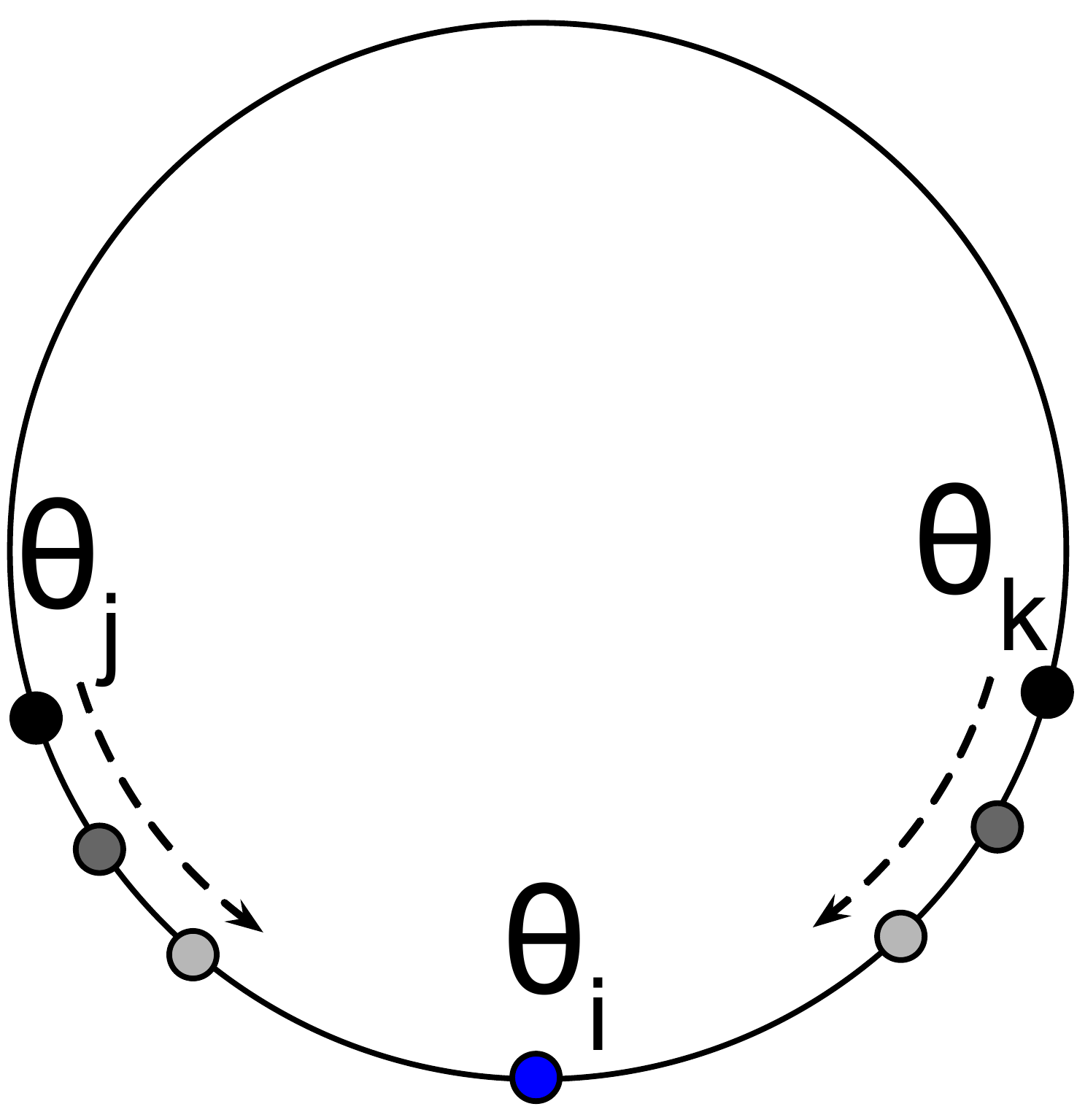}
	\subcaption{  Isosceles `breathers'}
	\label{f:breathers}
	\end{subfigure}	
	\caption{  In pendula, $\tht_i$ and $\tht_j$ form a molecule that along with $\tht_k$ oscillates about their common CM. In breathers, $\tht_i$ is at rest at the CM with $\tht_j$ and $\tht_k$ oscillating symmetrically about the CM. Here, $i,j$ and $k$ denote any permutation of the numerals $1$, $2$ and $3$.}
	\label{f:periodic-soln}
\end{figure}

Recall that the Euler and Lagrange solutions of the planar three-body problem arise through a reduction to the two body Kepler problem. We find an analogue of this construction for three rotors, where pendulum-like systems play the role of the Kepler problem. We find two such families of periodic orbits, the pendula and isosceles breathers (see Fig. \ref{f:periodic-soln}). They exist at all energies and go from librational to rotational motion as $E$ increases. They turn out to have remarkable stability properties which we deduce via their monodromy matrices.

\subsection{Periodic pendulum solutions} 
\label{s:pendulum-soln}

We seek solutions where one pair of rotors form a `bound state' with their angular separation remaining constant in time. We show that consistency requires this separation to vanish, so that the two behave like a single rotor and the equations reduce to that of a two-rotor problem. There are three such families of `pendulum' solutions depending on which pair is bound together (see Fig. \ref{f:pendula}). For definiteness, we suppose that the first two particles have a fixed separation $\zeta$ ($\tht_1 = \tht_2 + \zeta$ or $\vf_1 = \zeta$). Putting this in (\ref{e:3rotors-EOM-ph1ph2}), we get a consistency condition and an evolution equation for $\vf_2$:
	\beqs
	& 2 \sin \zeta - \sin \vf_2 + \sin(\zeta + \vf_2) = 0 \;\;  \text{and} \cr
	& m r^2 \ddot \vf_2 = - g \left[ 2 \sin \vf_2 - \sin \zeta + \sin(\zeta + \vf_2)\right].
	\eeqs
The consistency condition is satisfied only when the separation $\zeta = 0$, i.e., rotors 1 and 2 must coincide so that $\vf_1 = 0$ and $\dot \vf_1 = 0$ (or $p_2 = 2 p_1$) at all times (the other two families are defined by $\vf_2 = \dot \vf_2 = 0$ and $\vf_1+\vf_2 = \dot\vf_1 + \dot \vf_2 = 0$). The evolution equation for $\vf_2$ reduces to that for a simple pendulum: 
	\beq
	m r^2 \ddot \vf_2 =  - 3 g \sin \vf_2 \;\; \text{with} \;\; E = \fr{m r^2 \dot \vf_2^2}{3}  + 2g (1 - \cos \vf_2)
	\eeq
being the conserved energy. The periodic solutions are either librational (for $0 \leq E < 4g$) or rotational (for $E>4g$) and may be expressed in terms of the Jacobi elliptic function sn:  
	\beq
	\bar \vf_2(t) =
	\begin{cases}
	2 \arcsin (k \,{\rm sn} (\om_0 t, k))
	& \text{for} \quad 0 \leq E \leq 4g, \\
	 2 \arcsin ({\rm sn}(\om_0 t/\kappa, \kappa))	& \text{for} \quad E \geq 4g. 	\end{cases}
	\label{e:pendulum-solutions}
	\eeq
Here, $\om_0 = \sqrt{3g/mr^2}$ and the elliptic modulus $k=\sqrt{E/4g}$ with $\kappa = 1/k$. Thus $0 \leq k < 1$ for libration and $0 \leq \kappa < 1$ for rotation. The corresponding periods are $ \tau_{\rm lib} = 4 K(k)/\om_0$ and $\tau_{\rm rot} = 2\kappa K(\kappa)/\om_0$, where $K$ is the complete elliptic integral of the first kind. As $E \to 4g^\pm$, the period diverges and we have the separatrix $\bar \vf_2(t) =  2 \arcsin (\tanh (\om_0 t))$. The conditions $\vf_1 = 0$ and $p_2 = 2p_1$ define a 2d `pendulum submanifold' of the 4d phase space foliated by the above pendulum orbits. Upon including the CM motion of $\vf_0$, each of these periodic solutions may be promoted to a quasi-periodic orbit of the three-rotor problem. There is a two-parameter family of such orbits, labelled for instance, by the relative energy $E$ and the CM angular momentum $p_0$.

\subsubsection{Stability of pendulum solutions via monodromy matrix}
\label{s:pendulum-stability}

Introducing the dimensionless variables 
	\beq
	\tilde p_{1,2} = p_{1,2}/\sqrt{m r^2 g} \quad \text{and} \quad \tilde t =  t\sqrt{g/mr^2},
	\eeq 
the equations for small perturbations 
	\beq
	\vf_1 = \delta \vf_1, \;\; \vf_2 = \bar \vf_2 + \delta \vf_2 \;\; \text{and} \;\; p_{1 ,2} = \bar p_{1,2} + \delta p_{1,2}
	\eeq
to the above pendulum solutions (\ref{e:pendulum-solutions}) to (\ref{e:3rotors-CM-EOM-phasespace}) are
	\beq
	\fr{d^2}{d\tilde t^2} \colvec{2}{\delta \vf_1}{\delta \vf_2} = - \colvec{2}{2 + \cos \bar \vf_2 & 0}{\cos \bar \vf_2 - 1 & 3 \cos \bar \vf_2} \colvec{2}{\delta  \vf_1}{\delta \vf_2}.
	\eeq
This is a pair of coupled Lam\'e-type equations since $\bar \vf_2$ is an elliptic function. The analogous equation in the 2d anharmonic oscillator reduces to a single Lam\'e equation\cite{yoshida84, brack-mehta-tanaka}. Our case is a bit more involved and we will resort to a numerical approach here. To do so, it is convenient to consider the first order formulation
	\beq
	\fr{d}{d\tilde t}\colvec{4}{\delta  \vf_1}{\delta \vf_2}{\delta {\tilde{p}}_1}{\delta {\tilde{p}}_2} = -\colvec{4}{0 & 0 & -2 & 1}{0 & 0 & 1 & -2}{1 + \cos \bar \vf_2 & \cos \bar \vf_2 & 0 & 0}{\cos \bar \vf_2 & 2 \cos \bar \vf_2 & 0 & 0} \colvec{4}{\delta \vf_1}{\delta \vf_2}{\delta \tilde p_1}{\delta \tilde p_2}.
	\label{e:coeff-matrix}	
	\eeq
Since $m, g$ and $r$ have been scaled out, there is no loss of generality in working in units where $m = g = r = 1$, as we do in the rest of this section. The solution is $\psi(t) = U(t,0) \; \psi(0)$ where the real time-evolution matrix is given by a time-ordered exponential $U(t,0) = T\exp\{ \int_0^{t} A(t)\, dt\}$ where $A(t)$ is the coefficient matrix in (\ref{e:coeff-matrix}) and $T$ denotes time ordering. The tracelessness of $A(t)$ implies $\det U(t,0) = 1$ and preservation of phase volume. Though $A(t)$ is $\tau$-periodic, $\psi(t + \tau) = M(\tau) \psi(t)$ where the monodromy matrix $M(\tau) = U(t+\tau,t)$ is independent of $t$. Thus, $\psi(t + n \tau) = M^n \psi(t)$ for $n = 1,2,\ldots$, so that the long-time behavior of the perturbed solution may be determined by studying the spectrum of $M$. In fact, the eigenvalues $\la$ of $M$ may be related to the Lyapunov exponents associated to the pendulum solutions 
	\beq
	\mu = \lim_{t \to \infty} \fr{1}{t} \ln \fr{|\psi(t)|}{|\psi(0)|} \quad \text{via} \quad \mu = \frac{\log |\la|}{\tau}.
	\eeq
Since ours is a Hamiltonian system with 2 degrees of freedom, two of the eigenvalues of $M$ must equal one and the other two must be reciprocals \cite{hadjime}. On account of the reality of $M$, the latter two $(\la_3, \la_4)$ must be of the form $(e^{i \phi}, e^{- i \phi})$ or $(\la, 1/\la)$ where $\phi$ and $\la$ are real. It follows that two of the Lyapunov exponents must vanish while the other two must add up to zero. The stability of the pendulum orbit is governed by the stability index $\sigma = \tr M - 2$. We have stability if $|\sigma| \leq 2$ which corresponds to the eigenvalues $e^{\pm i \phi}$ and instability if $|\sigma| = |\la + 1/\la| > 2$.

\begin{figure*}	
	\centering
	\begin{subfigure}[t]{7cm}
		\centering
		\includegraphics[width=7cm]{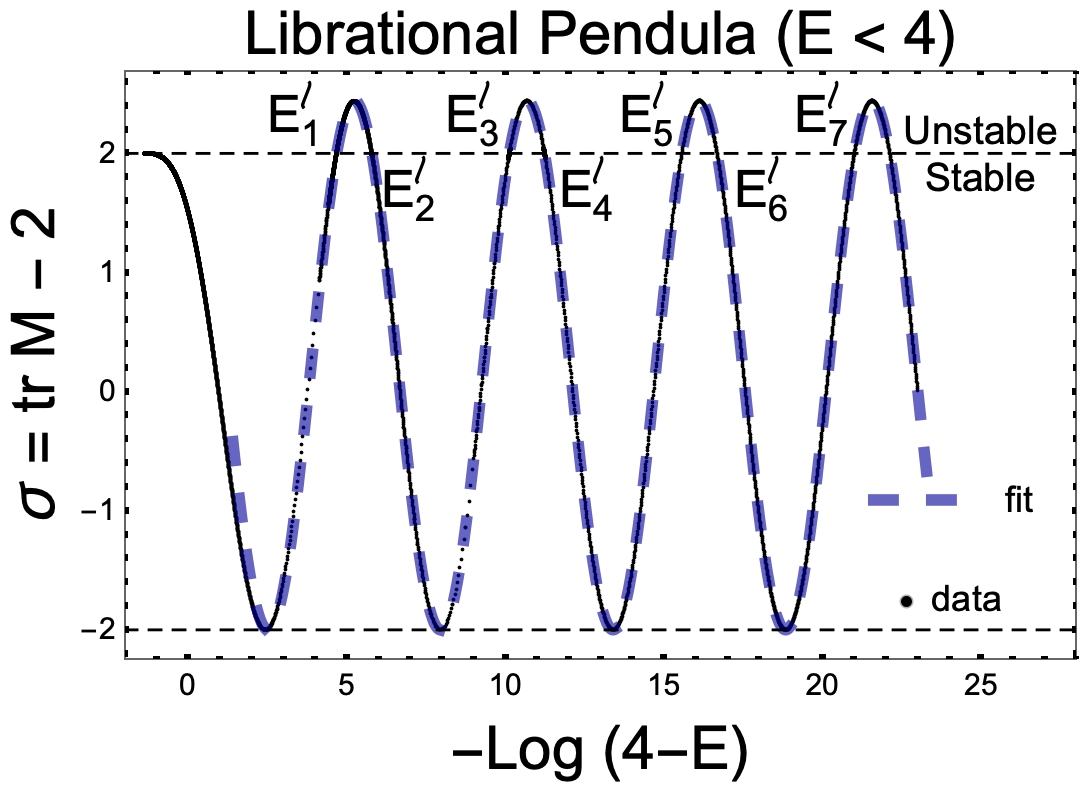}
	\end{subfigure}
\quad
	\begin{subfigure}[t]{7cm}
		\centering
		\includegraphics[width=7cm]{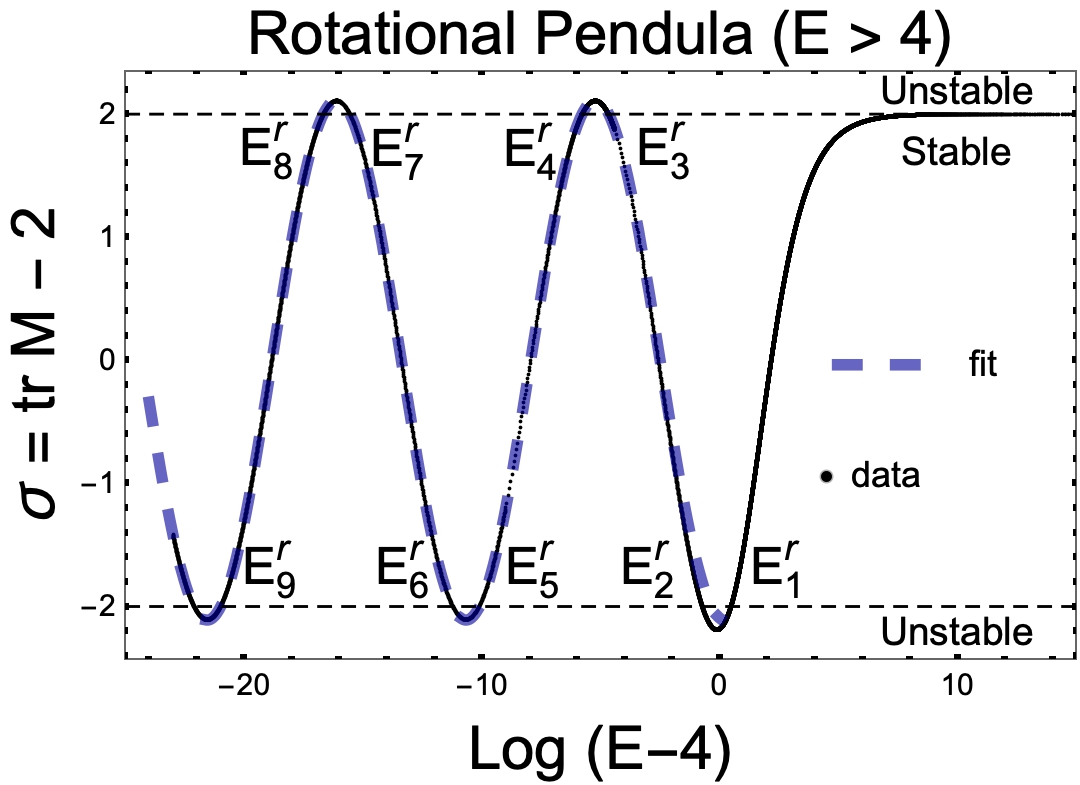}
	\end{subfigure}	
	\caption{  Numerically obtained stability index of pendulum solutions showing approach to periodic oscillations between stable and unstable phases as $E \to 4^\pm$. Equations (\ref{e:fit-trace-monodromy-lib}) and (\ref{e:fit-trace-monodromy-rot}) are seen to fit the data as $E \to 4^\pm$.}
	\label{f:monodromy-evals}
\end{figure*}

We now discuss the energy dependence of the stability index for pendula. In the limit of zero energy, (\ref{e:pendulum-solutions}) reduces to the ground state G and $A(t)$ becomes time-independent and similar to $2\pi i \times {\rm diag}(1, 1, -1, -1)$. Consequently, $M = \exp (A \tau)$ is the $4 \times 4$ identity $I$. Thus G is stable and small perturbations around it are periodic with period $\tau = 2\pi/\om_0$, as we know from Eq. (\ref{e:EOM-ph1ph2-perturbation-static-solution}). For $E > 0$, we evaluate $M$ numerically. We find it more efficient to regard $M$ as the fundamental matrix solution to $\dot \psi = A(t) \psi$ rather than as a path ordered exponential or as a product of infinitesimal time-evolution matrices. Remarkably, as discussed below, we find that while the system is stable for low energies $0 \leq E \leq E^\ell_1 \approx 3.99$ and high energies $E \geq E^\err_1 \approx 5.60$, the neighborhood of $E = 4$ consists of a doubly infinite sequence of intervals where the behavior alternates between stable and unstable (see Fig. \ref{f:monodromy-evals}). This is similar to the infinite sequence of transition energies for certain periodic orbits of a class of Hamiltonians studied in Ref.\cite{churchill} and to the singly infinite sequence of transitions in the 2d anharmonic oscillator as the coupling $\al$ goes from zero to infinity \cite{yoshida84}:
	\beq 
	H_{\rm anharm} = \half \left( p_1^2 + p_2^2 \right) + \ov{4} \left( q_1^4 + q_2^4 \right) + \al \: q_1^2 q_2^2.
	\label{e:anharmonic-oscillator-yoshida}
	\eeq
This accumulation of stable-to-unstable transition energies at the threshold for librational `bound' trajectories is also reminiscent of the quantum mechanical energy spectrum of Efimov trimers that form a geometric sequence accumulating at the zero-energy threshold corresponding to arbitrarily weak two-body bound states with diverging S-wave scattering length \cite{efimov-original}.

\subsubsection{Stability of librational pendula $(E < 4)$}

In the first stable phase $0 \leq E \leq E^\ell_1$, $\phi = \arg \la_3$ monotonically increases from $0$ to $2\pi$ with growing energy and $\la_4 = e^{-i \phi}$ goes round the unit circle once clockwise. There is a stable to unstable phase transition at $E_1^\ell$. In the unstable phase $E_1^\ell < E  < E_2^\ell$, $\sig > 2$ corresponding to  real positive $\la_4$ increasing from $1$ to $1.9$ and then dropping to $1$ (see Fig. \ref{f:monodromy-evals}). There is then an unstable to stable transition at $E^\ell_2$. This pattern repeats so that the librational regime $0 < E < 4$ is divided into an infinite succession of progressively narrower stable and unstable phases. Remarkably, we find that the stable phases asymptotically have equal widths on a logarithmic energy scale just as the unstable ones do. Indeed, if we let $E_{2n +1}^\ell$ and $E_{2n}^\ell$ denote the energies of the stable to unstable and unstable to stable transitions for $n = 1,2,3, \ldots$, then the widths $w^{lu}_n$ and $w^{ls}_{n+1}$ of the $n^{\rm th}$ unstable and $n+1^{\rm st}$ stable phases are 
	\beqs 
	w^{lu}_n &=& E^\ell_{2n} - E^\ell_{2n-1} \approx (E^\ell_{2} - E^\ell_{1}) \times e^{- \Lambda \, (n-1)} \quad \text{and} \cr
	w^{ls}_{n+1} &=& E^\ell_{2n+1} - E^\ell_{2n} \approx (E^\ell_{3} - E^\ell_{2}) \times e^{- \Lambda \, (n-1)}.
	\eeqs 
Here, $E^\ell_{2} - E^\ell_{1} \approx e^{-4.67} (1 - e^{-1.11})$ and $E^\ell_{3} - E^\ell_{2} \approx e^{-5.78} (1 - e^{-4.34})$ are the lengths of the first unstable and second stable intervals while $\Lambda \approx 1.11 + 4.34 = 5.45$ is the combined period on a log scale. The first stable phase has a width $E^\ell_1 - 0 \approx 4 - e^{-4.67}$ that does not scale like the rest. Our numerically obtained stability index (see Fig. \ref{f:monodromy-evals}) is well approximated by
	\beq
	\sigma \approx
	2.22 \cos \left[ \fr{2}{\sqrt{3}}\log (4 - E) + .24 \right] + .22 \; \text{as} \; E \to 4^-.
	\label{e:fit-trace-monodromy-lib}
	\eeq
On the other hand, $\sig(E) \sim 2 - {\cal O} (E^3)$ when $E \to 0$.

\subsubsection{Stability of rotational pendula $(E>4)$}

For sufficiently high energies $E \geq E_1^\err$, the rotational pendulum solutions are stable. In fact, as $E$ decreases from $\infty$ to $E^\err_1$, $\la_4 = e^{-i \phi}$ goes counterclockwise around the unit circle from 1 to $-1$. There is a stable to unstable transition at $E^\err_1$. As $E$ decreases from $E^\err_1$ to $E^\err_2$, $\la_4$ is real and negative, decreasing from $-1$ to $-1.5$ and then returning to $-1$ (see Fig. \ref{f:monodromy-evals}). This is followed by a stable phase for $E_2^\err \geq E \geq E_3^\err$ where $\la_4$ completes its passage counterclockwise around the unit circle reaching $1$ at $E_3^\err$. The last phase of this first cycle consists of an unstable phase between $E^\err_3$ and $E^\err_4$ where $\la_4$ is real and positive, increasing from $1$ to $1.4$ and then going down to $1$. The structure of this cycle is to be contrasted with those in the librational regime where $\la_4$ made complete revolutions around the unit circle in each stable phase and was always positive in unstable phases. This is reflected in the stability index overshooting both $2$ and $-2$ for rotational solutions but only exceeding 2 in the librational case. Furthermore, as in the librational case, there is an infinite sequence of alternating stable and unstable phases accumulating from above at $E = 4$, given by
	\beqs
	\text{stable energies} &=& \left [ E^\err_1 ,\infty \right ) \bigcup_{n = 1}^\infty \left[ E^\err_{2n+1}, E^\err_{2n} \right ] \cr
	\text{and unstable energies} &=& \bigcup_{n = 1}^\infty \left( E^\err_{2n}, E^\err_{2n-1} \right ).
	\eeqs
As before, with the exception of the two stable and one unstable intervals of highest energy, the widths of the stable and unstable energy intervals are approximately constant on a log scale:
	\beqs
	w^{ru}_n &=&  E^\err_{2n-1} - E^\err_{2n} \approx (E^\err_{3} - E^\err_{4}) \times e^{- \Lambda \, (n-2)} \quad \text{and} \cr
	w^{rs}_{n+1} &=& E^\err_{2n} - E^\err_{2n+1} \approx  (E^\err_{4} - E^\err_{5}) \times e^{- \Lambda \, (n-2)}
	\eeqs
for $n = 2,3,4 \cdots$. Here, $E^\err_{3} - E^\err_{4} \approx e^{-4.7} (1 - e^{-1.1})$ and $E^\err_{4} - E^\err_{5} \approx e^{-5.8} (1 - e^{-4.3})$ are the lengths of the second unstable and third stable intervals while $\Lambda \approx 1.1+4.3 = 5.4$ is the combined period. The three highest energy phases are anomalous: (a) $E \geq E_1^\err \approx 5.60$ is a stable phase of infinite width, (b) the unstable phase $E_1^\err > E > E_2^\err \approx 4.48$ has width $1.2 > 1.1$ on a log scale and manifests more acute instability and (c) the stable phase $E_2^\err \geq E \geq E_3^\err \approx 4.01$ has a less than typical width $3.9 < 4.3$ (see Fig. \ref{f:monodromy-evals}). As before, we obtain the fit
	\beq
	\sigma \approx - 2.11 \cos\left[ \fr{1}{\sqrt{3}} \log (E - 4) - .12 \right] \;\;  \text{as} \;\; E \to 4^+
	\label{e:fit-trace-monodromy-rot}
	\eeq
while $\sig(E) \sim 2- {\cal O} (1/E)$ when $E \to \infty$.
\subsubsection{Energy dependence of eigenvectors}

 Since the pendulum solutions form a one parameter family of periodic orbits $(0, \vf_2, p_1, 2p_1)$ with continuously varying time periods, a perturbation tangent to this family takes a pendulum trajectory to a neighboring pendulum trajectory and is therefore neutrally stable. These perturbations span the 1-eigenspace $\span (v_1,v_2)$ of the monodromy matrix, where $v_1 = (0, 1, 0, 0) = \pdr_{\vf_2}$ and $v_2 = (0, 0, 1, 2) = \pdr_{p_1} + 2 \pdr_{p_2}$. The other two eigenvectors of $M$ have a simple dependence on energy and thus help in ordering the eigenvalues $\la_3$ and $\la_4$ away from transitions. In the `unstable' energy intervals
	\beq
	( E^\ell_1, E^\ell_2 ) \cup ( E^\err_2, E^\err_1 ) \cup ( E^\ell_3, E^\ell_4 ) \cup ( E^\err_4, E^\err_3 ) \cup \ldots,
	\eeq 
$M = \text{diag}(1,1,\la_3, 1/\la_3)$ in the basis $(v_1, v_2, v_+, v_-)$ where $v_\pm = (2a(E), -a(E), \pm b (E), 0)$. In the same basis, $M = \text{diag} (1, 1, R_\phi)$ in the complementary `stable' energy intervals $( 0, E^\ell_1 ) \cup ( E^\err_1, \infty ) \cup \cdots$. Here, $R_\phi$ is the $2 \times 2$ rotation matrix $(\cos \phi, \sin \phi | - \sin \phi, \cos \phi)$. At the transition energies, either $a$ or $b$ vanishes so that $v_+$ and $v_-$ become collinear and continuity of eigenvectors with $E$ cannot be used to unambiguously order the corresponding eigenvalues across transitions. For instance, the eigenvalue that went counterclockwise around the unit circle for $E < E_1^\ell$ could be chosen to continue as the real eigenvalue of magnitude either greater or lesser than one when $E$ exceeds $E_1^\ell$.

\vspace{.2cm}

{\bf \fl Pitfall in trigonometric and quadratic approximation at low energies:} Interestingly, if for low energies ($0 \leq E \ll g$), we use the simple harmonic/trigonometric approximation to (\ref{e:pendulum-solutions}), $\bar \vf_2 \approx \sqrt{E/g} \sin \omega_0 t$ with $\om_0 = \sqrt{3g/mr^2}$ and $E \approx (mr^2/3) \dot {\bar \vf}_2^2 + g \bar\vf_2^2$ and approximate $\cos \bar \vf_2$ by $1 - \bar\vf_2^2/2$ in (\ref{e:coeff-matrix}), we find that the eigenvalues of the monodromy matrix are of the form $e^{\pm i \tht}$ and $e^{\pm i \phi}$ where $\tht$ and $\phi$ monotonically increase from zero with energy upto moderate energies. By contrast, as we saw above, two of the eigenvalues $\la_{1,2}$ are always equal to one, a fact which is not captured by this approximation. 

\subsection{Periodic isosceles `breather' solutions}
\label{s:iso-periodic-sol}

We seek solutions where two of the separations remain equal at all times: $\tht_i - \tht_j = \tht_j - \tht_k$ where ($i, j, k$) is any permutation of (1,2,3). Loosely, these are `breathers' where one rotor is always at rest midway between the other two (see Fig. \ref{f:breathers}). For definiteness, suppose $\tht_1 - \tht_2 = \tht_2 - \tht_3$ or equivalently $\vf_1 = \vf_2$. Putting this  in Eq. (\ref{e:3rotors-EOM-ph1ph2}), we get a single evolution equation for $\vf_1 = \vf_2 = \vf$,
	\beq
	m r^2 \ddot \vf = - g (\sin \vf + \sin 2\vf),
	\label{e:isosceles-EOM}
	\eeq
which may be interpreted as a simple pendulum with an additional periodic force. As before, each periodic solution of this equation may, upon inclusion of CM motion, be used to obtain quasi-periodic solutions of the three-rotor problem.

  \begin{figure}
  \includegraphics[width=6cm]{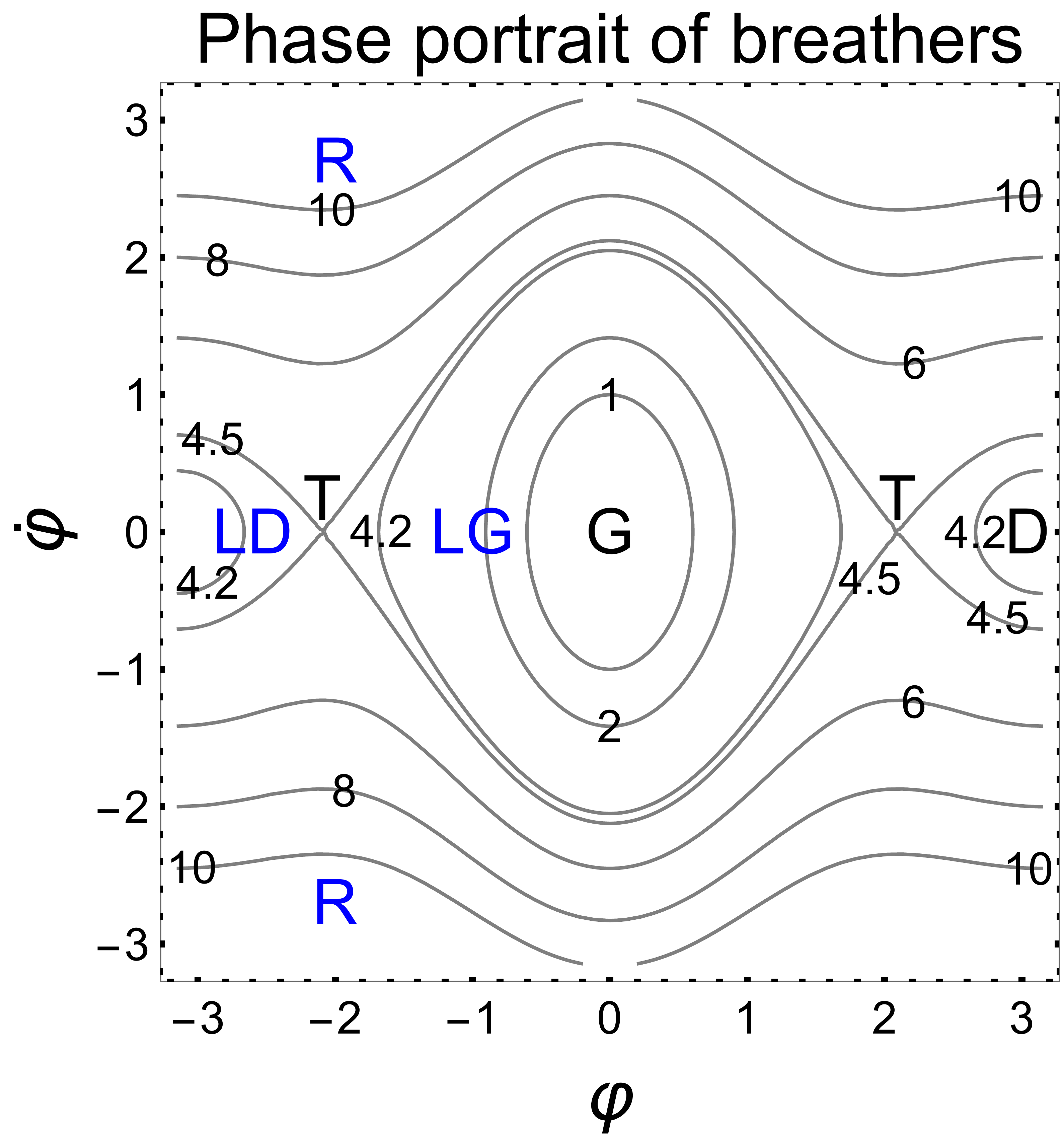}
  \caption{\label{f:isosceles-phaseportrait} Level contours of $E$ on a phase portrait of the LG, LD and R families of isosceles periodic solutions.}
  \end{figure}

 At $E = 0$, the isosceles solutions reduce to the ground state G. More generally, there are two families of librational breathers. With $E$ denoting energy in units of $g$, they are LG (oscillations around G $(\vf = 0)$ for $0 \leq E \leq 9/2$) and LD (oscillations around D $(\vf = \pi)$ for $4 \leq E \leq 9/2$) with monotonically growing time period which diverges at the separatrix at $E = 9/2$ (see Fig.~\ref{f:isosceles-phaseportrait}). For $E > 9/2$, we have rotational modes R with time period diminishing with energy ($\tau^{\rm rot}(E) \sim 2\pi/\sqrt{E}$ as $E \to \infty$). At very high energies, one rotor is at rest while the other two rotate rapidly in opposite directions. Eq. (\ref{e:isosceles-EOM}) may be reduced to quadrature by use of the conserved relative energy (\ref{e:egy-3rotors-phi1-phi2-coords}):
	\beq
	E = m r^2 \dot \vf^2 + g (3 - 2 \cos \vf - \cos 2 \vf).
	\eeq
For instance, in the case of the LG family,
	\beq
	\fr{\om_0 t}{\sqrt{3}} 
	= \fr{1}{\sqrt{2}} \int_0^{u} \fr{du}{\sqrt{u(2-u)(u^2 - 3u + E/2)}}
	\eeq
where $u = 1- \cos \vf$. The relative angle $\vf$ may be expressed in terms of Jacobi elliptic functions. Putting $\eps = \sqrt{9-2E}$,
	\beqs
	\vf(t) &=& \arccos \left( 1- \fr{E \eta^2}{2 \eps + (3-\eps) \eta^2} \right) \quad \text{where} \cr
	 \eta(t) &=& \text{sn} \left(\fr{\sqrt{\eps}\om_0 t}{\sqrt{3}} , \sqrt{\fr{(\eps - 1)(3 - \eps)}{8 \eps}} \right).
	\eeqs
It turns out that the periods of both LG and LD families are given by a common expression,
	\beq
	\tau^{\rm lib}(E) = \fr{4\sqrt{3}}{\om_0 \sqrt{\eps}} K\left( \sqrt{\ov{2} - \fr{6 - E}{4 \eps}} \right) \;\; \text{for} \;\; 0 \leq E \leq 4.5.
	\eeq
As $E \to 4.5$, $\tau^{\rm lib}$ diverges as $2 \sqrt{2/3} \log (4.5 - E)$. The time period of rotational solutions (for $E \geq 4.5$) is
	\beq
	\tau^{\rm rot}(E) =\fr{4\sqrt{3}}{\om_0} (E^2 - 4 E)^{-1/4} K \left(\sqrt{ \half + \fr{6 - E}{2\sqrt{E^2 - 4E}}} \right).
	\eeq

\subsubsection{Linear stability of breathers}
\label{s:stability-of-breathers}

\begin{figure}			\includegraphics[width=6cm]{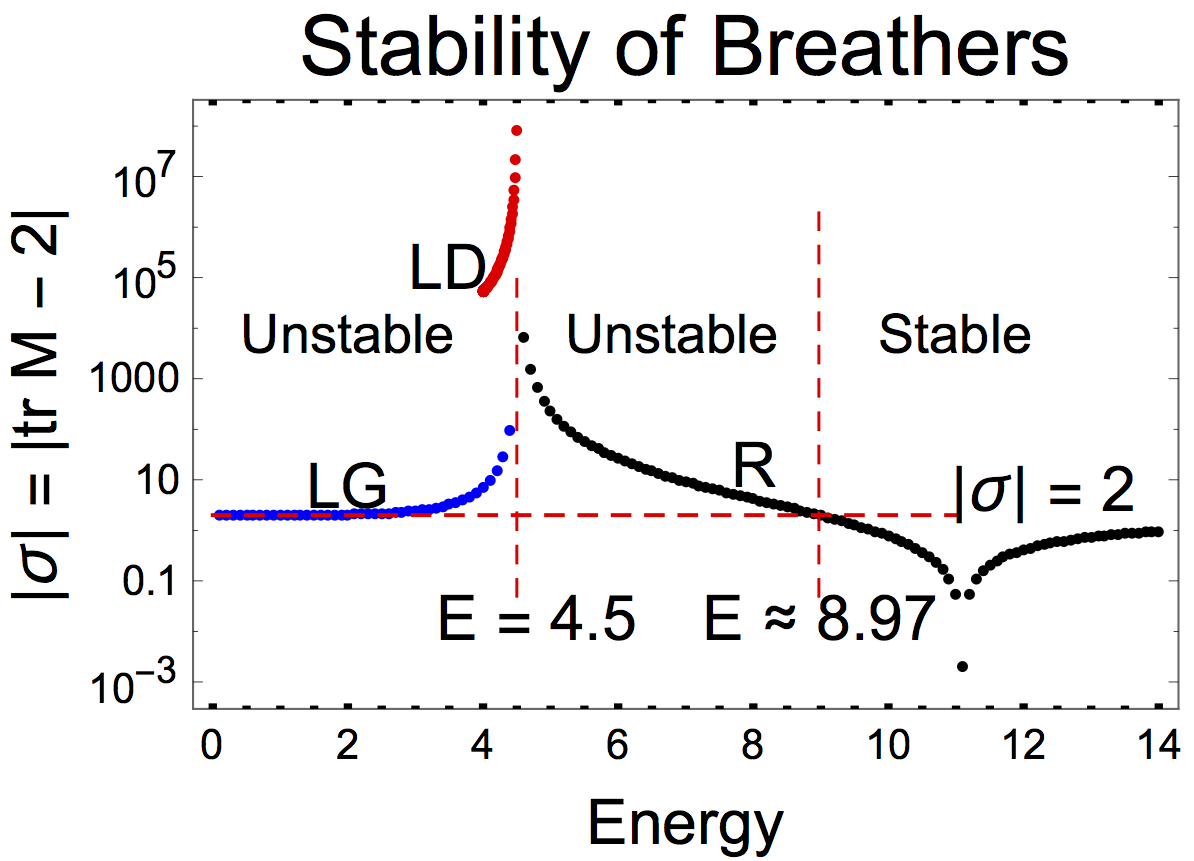}
		\caption{\label{f:isosceles-stability-index} Absolute value of the stability index of the isosceles breathers as a function of energy.}
\end{figure}

The stability of isosceles solutions as encoded in the stability index $(\sigma = \tr M -2)$ is qualitatively different from that of the pendulum solutions. In particular, there is only one unstable to stable transition occuring at $E \approx 8.97$ (see Fig. \ref{f:isosceles-stability-index}). Indeed, by computing the monodromies, we find that both families LG and LD of librational solutions are unstable. The stability index $\sig_{\rm LG}$ grows monotonically from $2$ to $\infty$ as the energy increases from $0$ to $4.5$. In particular, even though arbitrarily low energy breathers are small oscillations around the stable ground state G, they are themselves unstable to small perturbations. By contrast, we recall that low energy pendulum solutions around G are stable. On the other hand, the LD family of breathers are much more unstable, indeed, we find that $\sig_{\rm LD}$ increases from $\approx 5.3\times10^4$ to $\infty$ for $4 < E < 4.5$. This is perhaps not unexpected, given that they are oscillations around the unstable static solution D. The rotational breathers are unstable for $4.5 < E < 8.97$ with $\sig_{\rm R}$ growing from $-\infty$ to $-2$. These divergences of $\sigma$ indicate that isosceles solutions around $E = 4.5$ suffer severe instabilities not seen in the pendulum solutions. Beyond $E = 8.97$, the rotational breathers are stable with $\sig_{\rm R}$ growing from $-2$ to $2$ as $E \to \infty$. This stability of the breathers is also evident from the Poincar\'e sections of \S \ref{s:poincare-section}. In fact, the isosceles solutions go from intersecting the Poincar\'e surface `$\vf_1 = 0$' at  hyperbolic to elliptic fixed points as the energy is increased beyond $E \approx 8.97$ (see Fig. \ref{f:psec-egy=2-3}-\ref{f:psec-egy=4.5-to-18}).

\section{Jacobi-Maupertuis metric and curvature}
\label{s:JM-approach}

We now consider a geometric reformulation of the classical three-rotor problem, that suggests the emergence of widespread instabilities for $E > 4$ from a largely stable phase at lower energies and a return to regularity as $E \to \infty$.  This indicates the presence of an `order-chaos-order' transition which will be confirmed in \S \ref{s:poincare-section}.

It is well known that configuration space trajectories of the Lagrangian $L = (1/2) m_{ij} \dot q_i \dot q_j - V(q)$ may be regarded as reparametrized geodesics of the Jacobi-Maupertuis (JM) metric $g^{\rm JM}_{ij} = (E - V) m_{ij}$ which is conformal to the mass/kinetic metric $m_{ij}(q)$ \cite{lanczos, gskhs-three-body}. The sectional curvatures of this metric have information on the behavior of nearby trajectories with positive/negative curvature associated to (linear) stability/instability. For the three-rotor problem, the JM metric on the $\vf_1$-$\vf_2$ configuration torus is given by
	\beq
	ds_{\rm JM}^2 = \fr{2 m r^2}{3} (E - V) (d\vf_1^2 + d\vf_1 d\vf_2 + d\vf_2^2),
	\eeq
where $V = g[3 - \cos \vf_1 - \cos \vf_2 - \cos( \vf_1 + \vf_2) ]$. Letting $f$ denote the conformal factor $E - V$ and using the gradient and Laplacian defined with respect to the flat kinetic metric, the corresponding scalar curvature ($2 \times$ the Gaussian curvature) is
	\beqs
	&R = \fr{|\grad f|^2 - f \Delta f}{f^3} = \fr{g^2}{m r^2 (E-V)^3} \bigg[6 + (\fr{2E}{g} - 3)(3 - \fr{V}{g}) + \cr
	& \cos(\vf_1-\vf_2)   + \cos(2\vf_1+\vf_2) + \cos(\vf_1+2\vf_2) \bigg].
 \label{e:JM-scalar-curv}
	\eeqs

\subsection{Behavior of JM curvature}

\begin{figure*}	
	\centering
	\begin{subfigure}[t]{2.9cm}
		\centering
		\includegraphics[width=2.9cm]{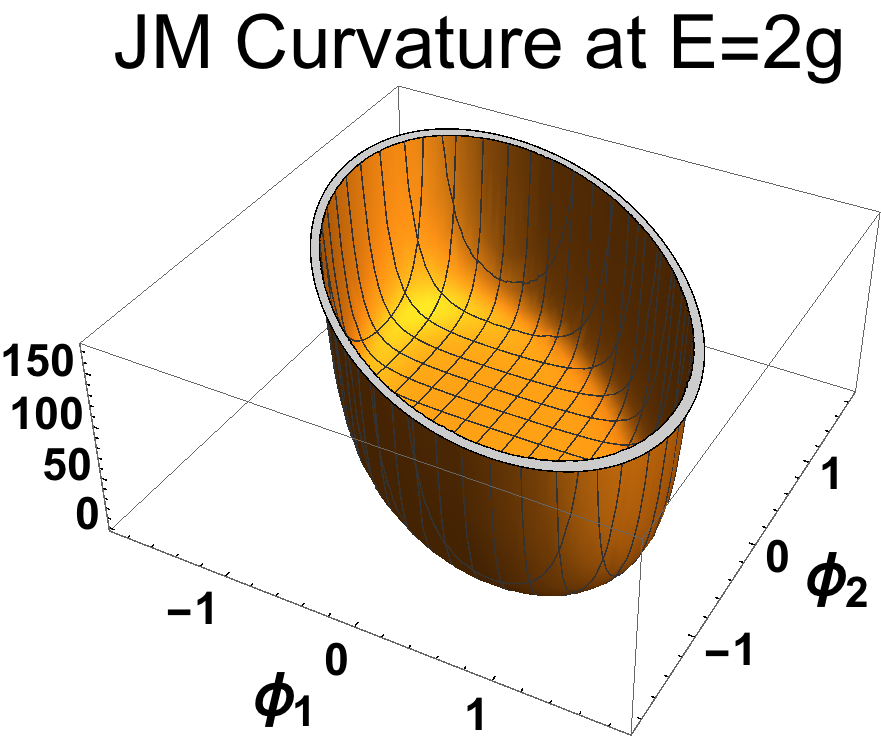}
	\end{subfigure}	
	\qquad
	\begin{subfigure}[t]{2.9cm}
		\centering
		\includegraphics[width=2.9cm]{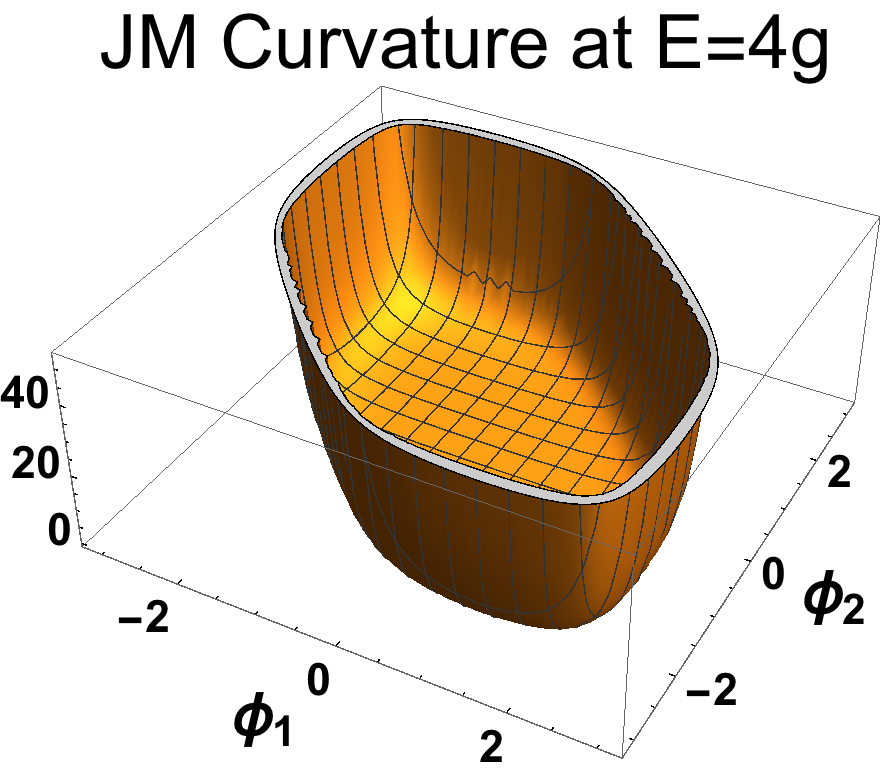}
	\end{subfigure}	
	\qquad
	\begin{subfigure}[t]{2.8cm}
		\centering
		\includegraphics[width=2.8cm]{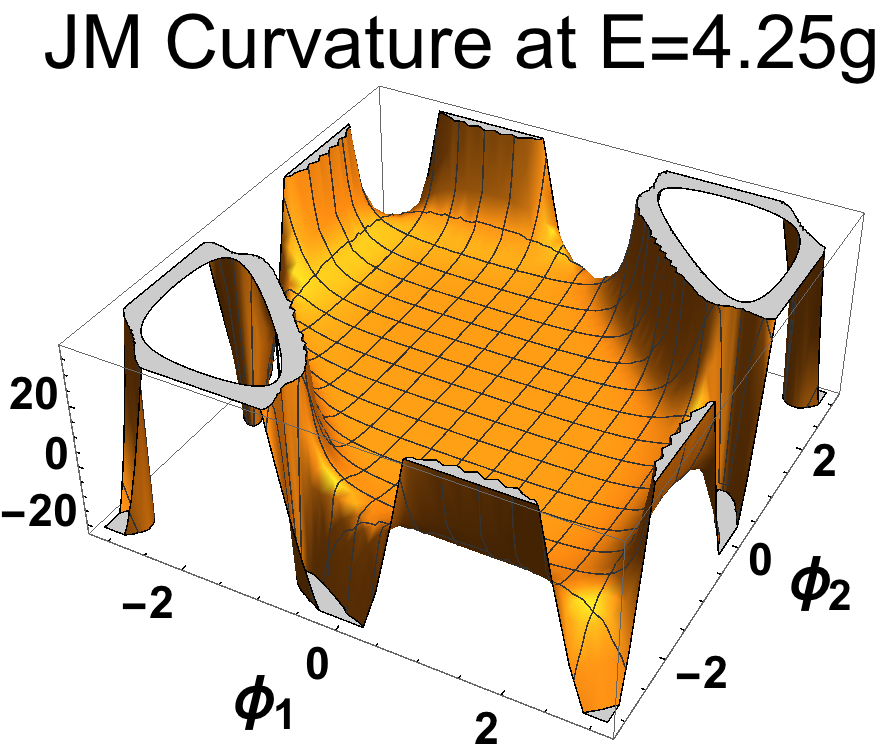}
	\end{subfigure}	
	\qquad
	\begin{subfigure}[t]{2.8cm}
		\centering
		\includegraphics[width=2.8cm]{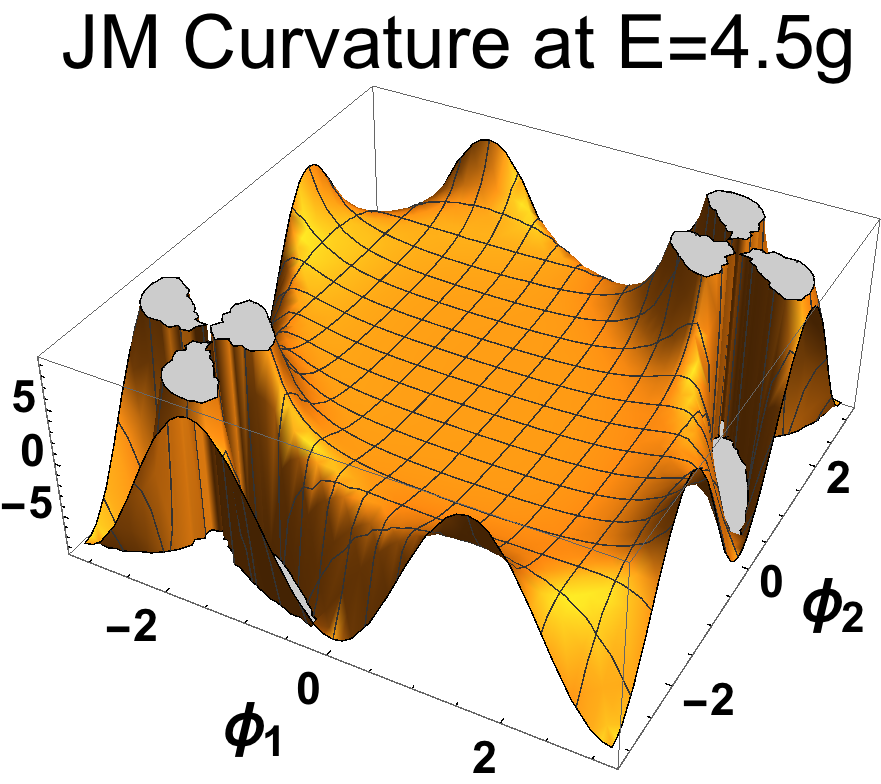}
	\end{subfigure}
	\qquad
	\begin{subfigure}[t]{2.8cm}
		\centering
		\includegraphics[width=2.8cm]{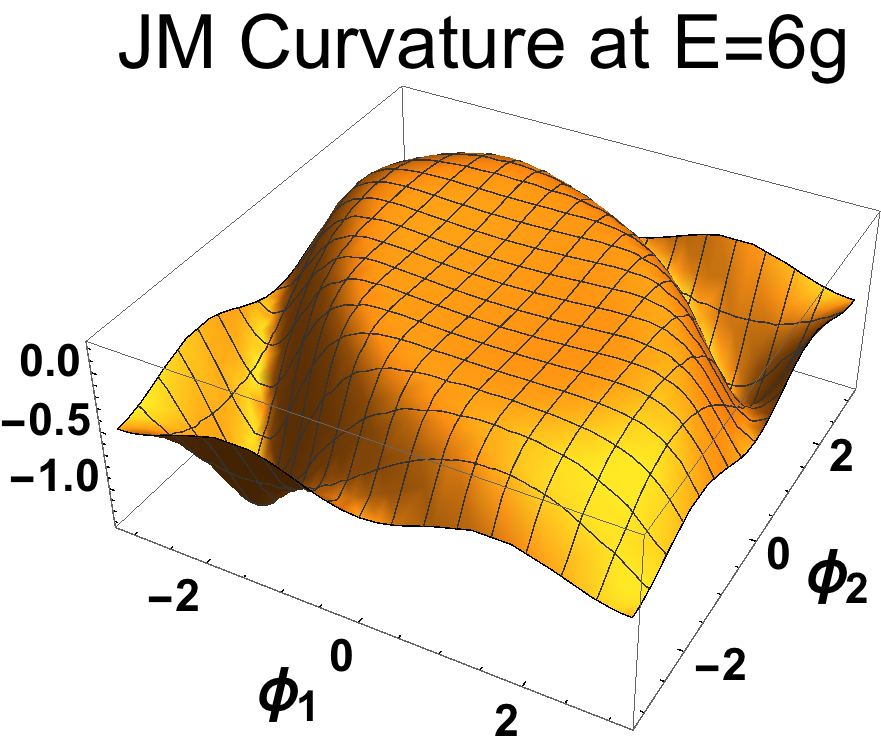}
	\end{subfigure}	
	\caption{  Scalar curvature $R$ of the JM metric on the Hill region of the $\vf_1$-$\vf_2$ torus. In the regions shaded grey, $|R|$ is very large. We see that $R > 0$ for $E \leq 4g$ but has both signs for $E > 4g$ indicating instabilities.}
	\label{f:curvature-3rotors-phi1-phi2-torus}
\end{figure*}

 For $0 \leq E \leq 4g$, $R$ is strictly positive  in the classically allowed Hill region $(V < E)$ and diverges on the Hill boundary $V = E$ where the conformal factor vanishes (see Appendix \ref{a:positivity-of-JM-curvature} for a proof and the first two `bath-tub' plots of $R$ in Fig.~\ref{f:curvature-3rotors-phi1-phi2-torus}). Thus the geodesic flow should be stable for these energies. Remarkably, we also find a near absence of chaos in all Poincar\'e sections for $E \lesssim 3.8g$ (see Fig. \ref{f:psec-egy=2-3} and  \ref{f:chaos-vs-egy}). We will see that Poincar\'e surfaces show significant chaotic regions for $E > 4g$. This is perhaps related to the instabilities associated with $R$ acquiring both signs above this energy. Indeed, for $4g < E \leq 9g/2$, the above `bath-tub' develops sinks around the saddles $D(0,\pi)$, $D(\pi,0)$ and $D(\pi,\pi)$  where $R$ becomes negative, though it continues to diverge on the Hill boundary which is a union of two closed curves encircling the local maxima T$(\pm 2\pi/3, \pm 2\pi/3)$. For $E > 9g/2$, the Hill region expands to cover the whole torus. Here, though bounded, $R$ takes either sign while ensuring that the total curvature $\int_{T^2} R \, \sqrt{\det g_{ij}}\; d\vf_1 \, d\vf_2$ vanishes. For asymptotically high energies, the JM metric tends to the flat metric $E \,m_{ij}$ and $R \sim 1/E^2 \to 0$ everywhere indicating a return to regularity.

\subsection{JM stability of static solutions}

 The static solutions G, D and T lie on the boundary of the Hill regions corresponding to the energies $E_{\rm G, D, T} = 0$, $4g$ and $4.5 g$. We define the curvatures at G, D and T by letting $E$ approach the appropriate limiting values in the following formulae: 
	\beqs
	&R_{(0,0)} = \fr{6 g}{m r^2 E^2}, \quad R_{(0,\pi),(\pi,0),(\pi,\pi)} = \fr{- 2 g/ m r^2}{(E - 4g)^2} \cr
	 &\text{and} \quad R_{\left(\pm \fr{2\pi}{3},\pm \fr{2\pi}{3}\right)} = \fr{- 12 g/m r^2}{(2E -9g)^2}.
	\eeqs
Thus $R_{\rm G} = \infty$ while $R_{\rm D} = R_{\rm T} = - \infty$ indicating that G is stable while D and T are unstable. These results on geodesic stability are similar to those obtained from (\ref{e:EOM-ph1ph2-perturbation-static-solution}). Note that we do not define the curvatures at G, D and T by approaching these points from within the Hill regions as these limits are not defined for G and T and gives $+\infty$  for D. On the other hand, it is physically forbidden to approach the Hill boundary from the outside. Thus we approach G, D and T by varying the energy while holding the location on the torus fixed.

\section{Poincar\'e sections: periodic orbits and chaos}
\label{s:poincare-section}

To study the transitions from integrability to chaos in the 3-rotor problem, we use the method of Poincar\'e sections. Phase trajectories are constrained to lie on energy level sets which are compact 3d submanifolds of the 4d phase space parametrized by $\vf_1$, $\vf_2$, $p_1$ and $p_2$ (cotangent bundle of the 2-torus). By the Poincar\'e surface `$\vf_1 = 0$' at energy $E$ (in units of $g$), we mean the 2d surface $\vf_1 = 0$ contained in a level-manifold of energy. It may be parametrized by $\vf_2$ and $p_2$ with the two possible values of $p_1(\vf_2,p_2;E)$ determined by the conservation of energy. Similarly, we may consider other Poincar\'e surfaces such as the ones defined by $\vf_2 = 0$, $p_1 = 0$, $p_2 = 0$ etc. We record the points on the Poincar\'e surface where a trajectory that begins on it returns to it under the Poincar\'e return map, thus obtaining a Poincar\'e section for the given initial condition (IC). For transversal intersections, a periodic trajectory leads to a Poincar\'e section consisting of finitely many points while quasi-periodic trajectories produce sections supported on a finite union of 1d curves. We call these two types of sections `regular'. By a chaotic section, we mean one that is not supported on such curves but explores a 2d region. In practice, deciding whether a numerically obtained Poincar\'e section is regular or chaotic can be a bit ambiguous in borderline cases when it is supported on a thickened curve (see Fig. \ref{f:psec-egy=18} and around $I$ in Fig. \ref{f:psec-egy=2-3}). We define the chaotic region of a Poincar\'e surface at energy $E$ to be the union of all chaotic sections at that energy.

\subsection{Transition to chaos and global chaos}
\label{s:transition-to-chaos-global-chaos}

\begin{figure*}
	\centering
	\begin{subfigure}[t]{4.5cm}
		\centering
		\includegraphics[width=4.5cm]{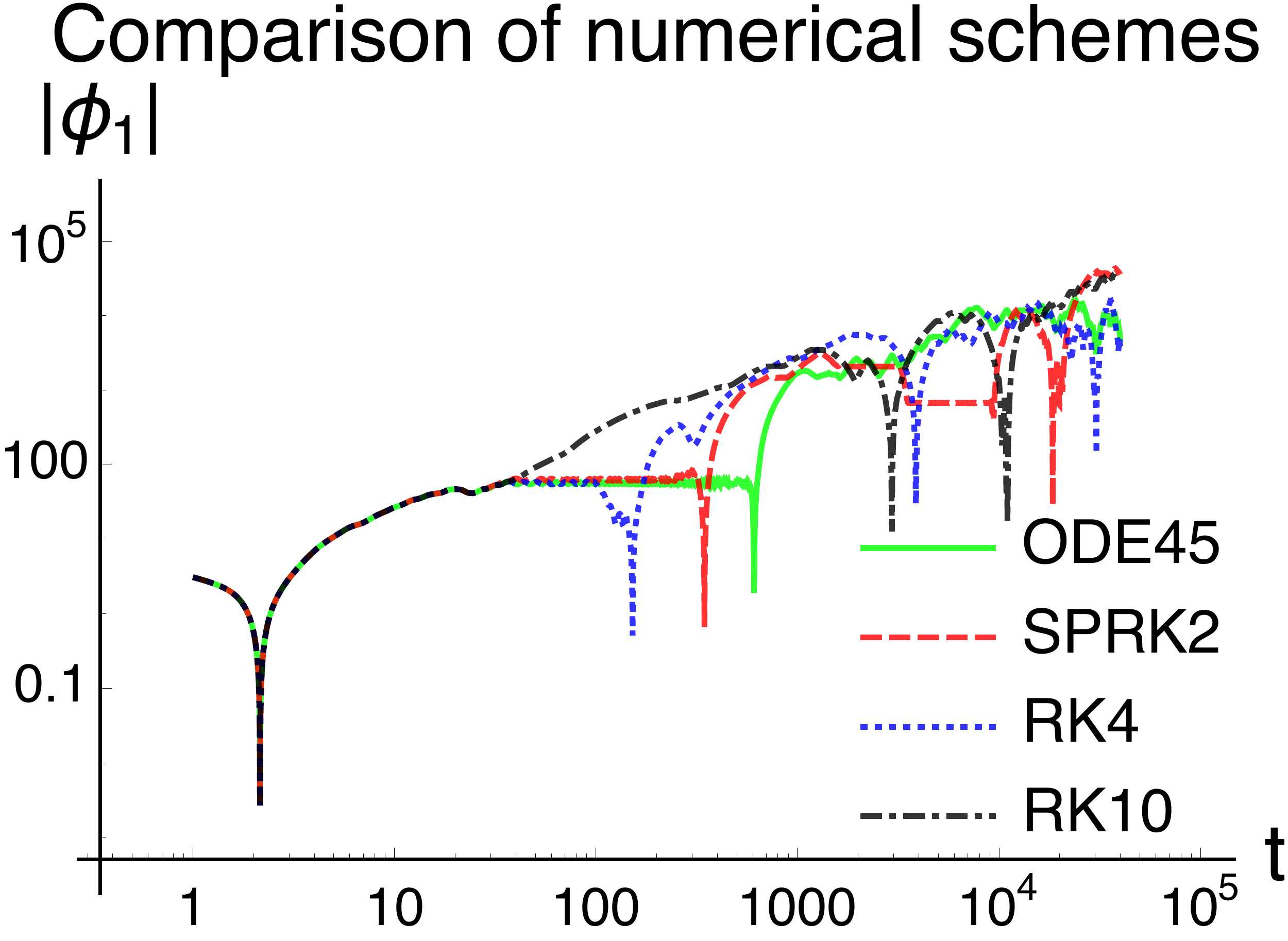}
		\caption{}
	\end{subfigure}
\quad	
	\begin{subfigure}[t]{3cm}
		\centering
		\includegraphics[width=3cm]{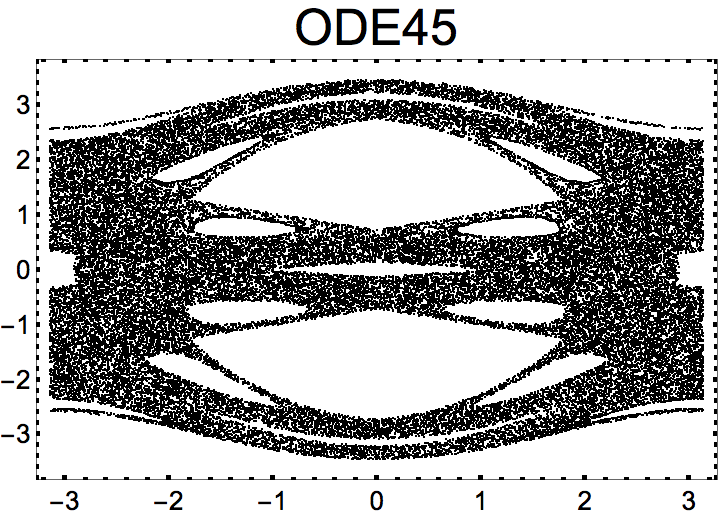}
		\caption{}
	\end{subfigure}
	\begin{subfigure}[t]{3cm}
		\centering
		\includegraphics[width=3cm]{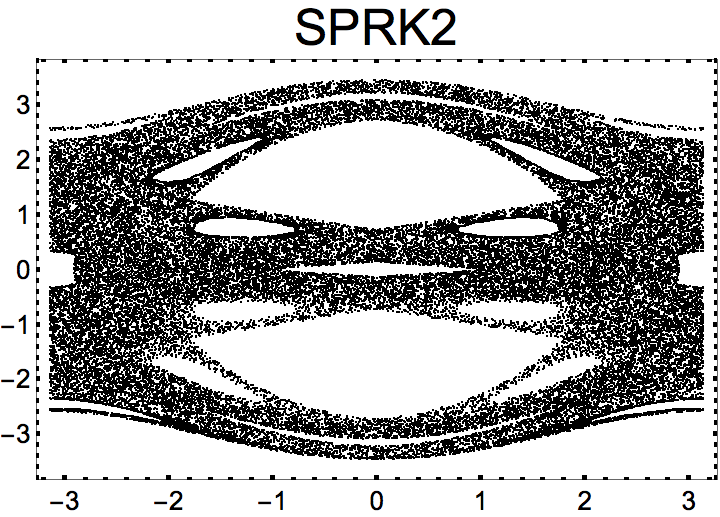}
		\caption{}
	\end{subfigure}
	\begin{subfigure}[t]{3cm}
		\centering
		\includegraphics[width=3cm]{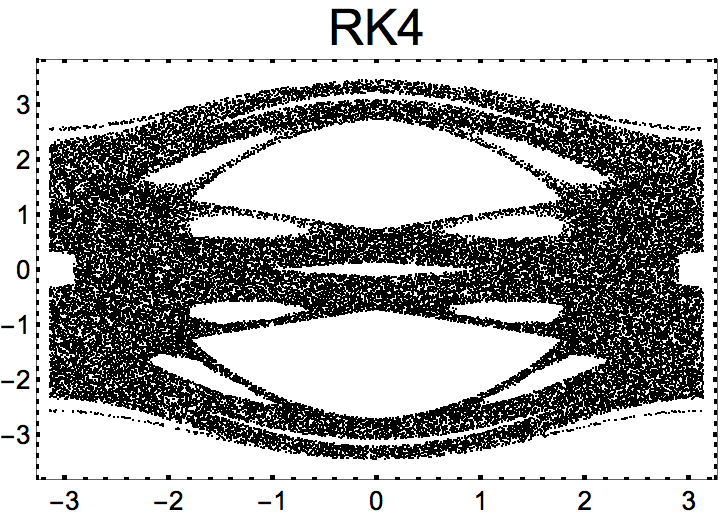}
		\caption{}
	\end{subfigure}
	\begin{subfigure}[t]{3cm}
		\centering
		\includegraphics[width=3cm]{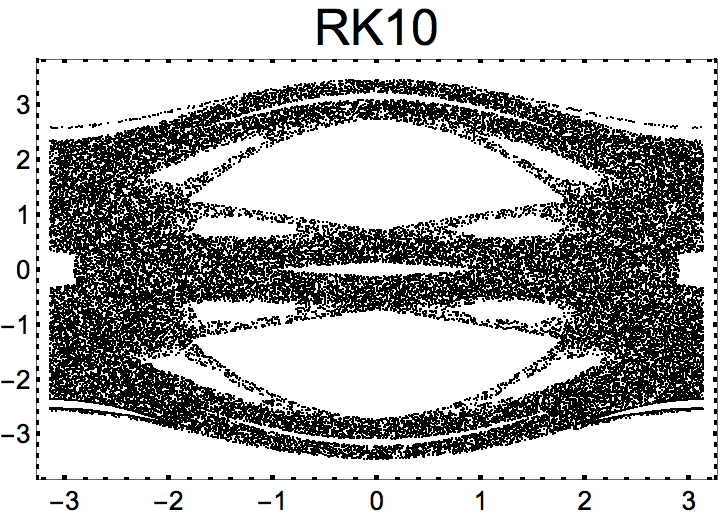}
		\caption{}
	\end{subfigure}
	\caption{  (a) The trajectories (e.g., $|\vf_{1}|$) obtained via different numerical schemes cease to agree after $t \sim 10^2$ for the IC $\vf_1 = 6.23$, $\vf_2 = 3.00$, $p_1 = -.90$ and $p_2 = 1.87$ with $E=9.98$. (b, c, d, e) However, Poincar\'e sections (with $\approx 5 \times 10^4$ points) obtained via different schemes are seen to explore qualitatively similar regions when evolved till $t = 10^5$ (though {\em not} for shorter times $\sim 10^3$).}	
	\label{f:scheme-dependent-trajectories}
\end{figure*}

\subsubsection{Numerical schemes and robustness of Poincar\'e sections}

 To obtain Poincar\'e sections, we implement the following numerical schemes: ODE45: explicit Runge-Kutta with difference order 5; RK4 and RK10: implicit Runge-Kutta with difference orders 4 and 10 and SPRK2: symplectic partitioned Runge-Kutta with difference order 2. Due to the accumulation of errors, different numerical schemes (for the same ICs) sometimes produce trajectories that cease to agree after some time, thus reflecting the sensitivity to initial conditions. Despite this difference in trajectories, we find that the corresponding Poincar\'e sections from all schemes are roughly the same when evolved for sufficiently long times (see Fig.~\ref{f:scheme-dependent-trajectories}). Moreover, we find a strong correlation between the degree to which different schemes produce the same trajectory and the degree of chaos as manifested in Poincar\'e sections. As the agreement in trajectories between different schemes improves, the Poincar\'e sections go from being spread over 2d regions to being concentrated on a finite union of 1d curves. Since ODE45 is computationally faster than the other schemes, the results presented below are obtained using it. Furthermore, we find that for all ICs studied, all four Poincar\'e sections on surfaces defined by $\vf_1 = 0$, $\vf_2 = 0$, $p_1 = 0$ and $p_2 = 0$ are qualitatively similar with regard to the degree of regularity or chaos. Thus, in the sequel, we restrict to the Poincar\'e surface defined by $\vf_1 = 0$. 

\subsubsection{Symmetry breaking accompanying  onset of chaos}

\begin{figure}	
	\centering
	\begin{subfigure}[t]{4.1cm}
		\centering
		\includegraphics[width=4.1cm]{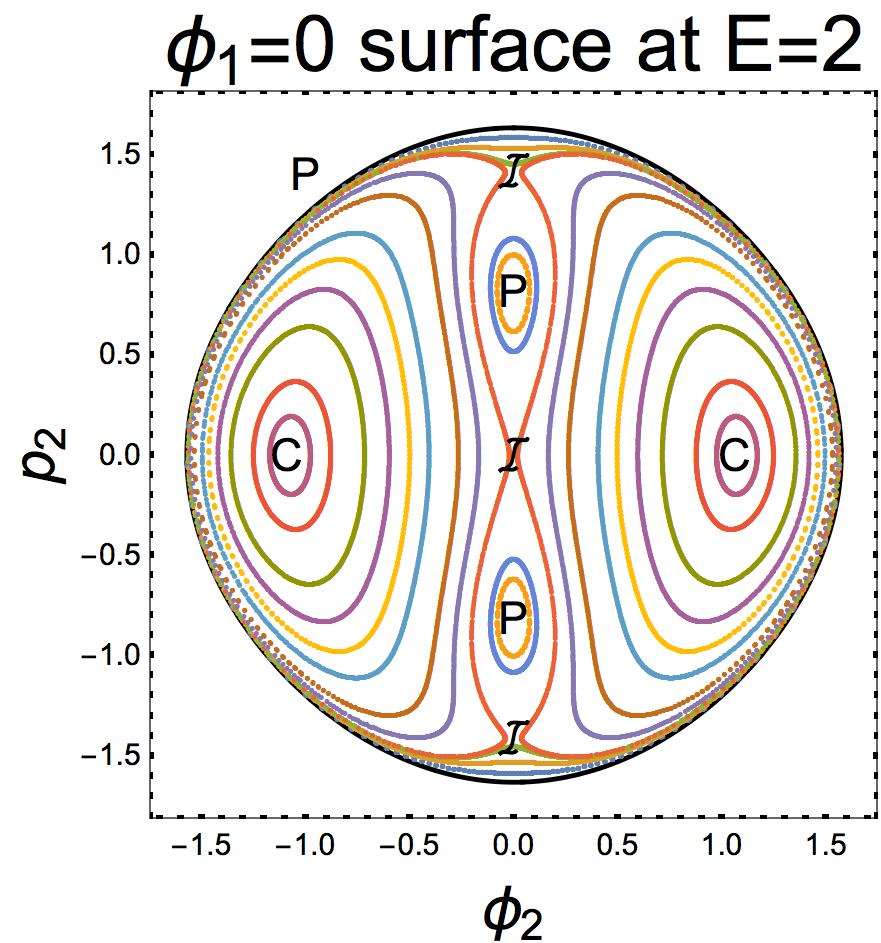}
	\end{subfigure}	
\;\;
	\begin{subfigure}[t]{4.1cm}
		\centering
		\includegraphics[width=4.1cm]{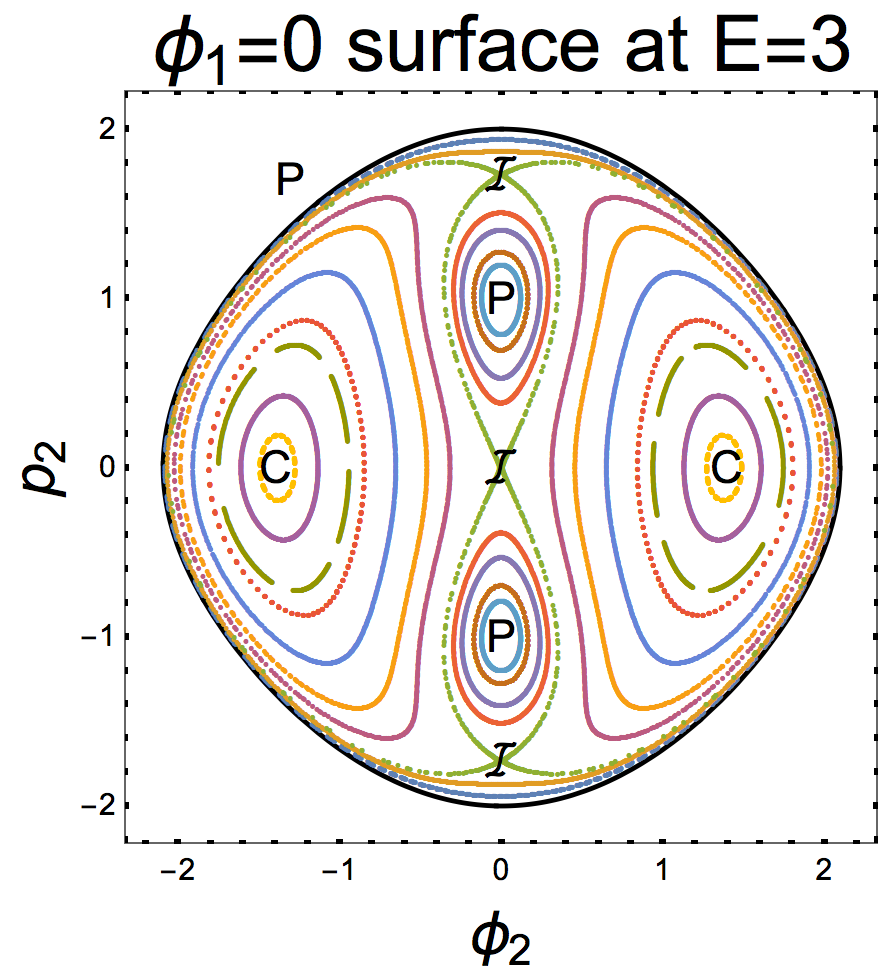}
	\end{subfigure}	
	\caption{  Several Poincar\'e sections in the energetically allowed `Hill' region on the `$\vf_1 = 0$' surface for $E = 2$ and $3$. All sections (indicated by distinct colors online) are largely regular and possess up-down and left-right symmetries. The Hill boundary is the librational pendulum solution $\vf_1 = 0$. P, ${\cal I}$ and C indicate pendulum, isosceles and choreography periodic solutions. More careful examination of the vicinity of the $\cal I$s shows small chaotic sections.}
	\label{f:psec-egy=2-3}
\end{figure}

\begin{figure*}
	\begin{subfigure}[t]{5.7cm}
		\includegraphics[width=5.7cm]{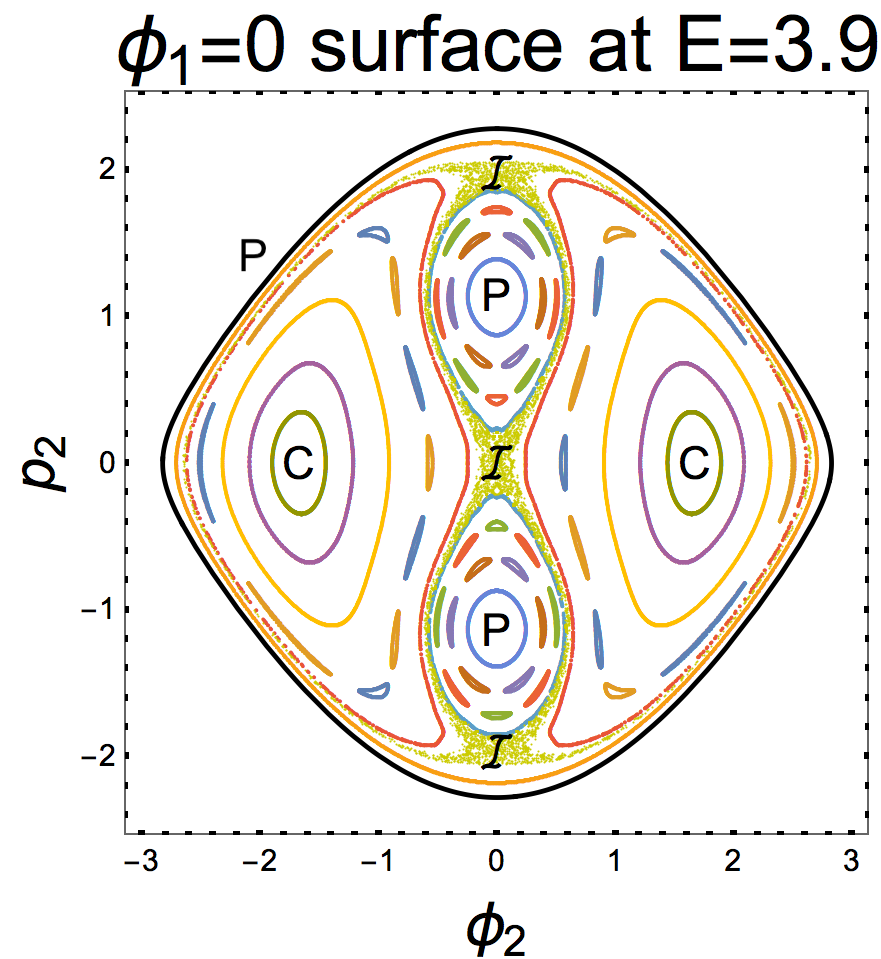}
		\caption{}
		\label{f:psec-egy=3.9}
	\end{subfigure}
\quad	
	\begin{subfigure}[t]{5.7cm}
		\includegraphics[width=5.7cm]{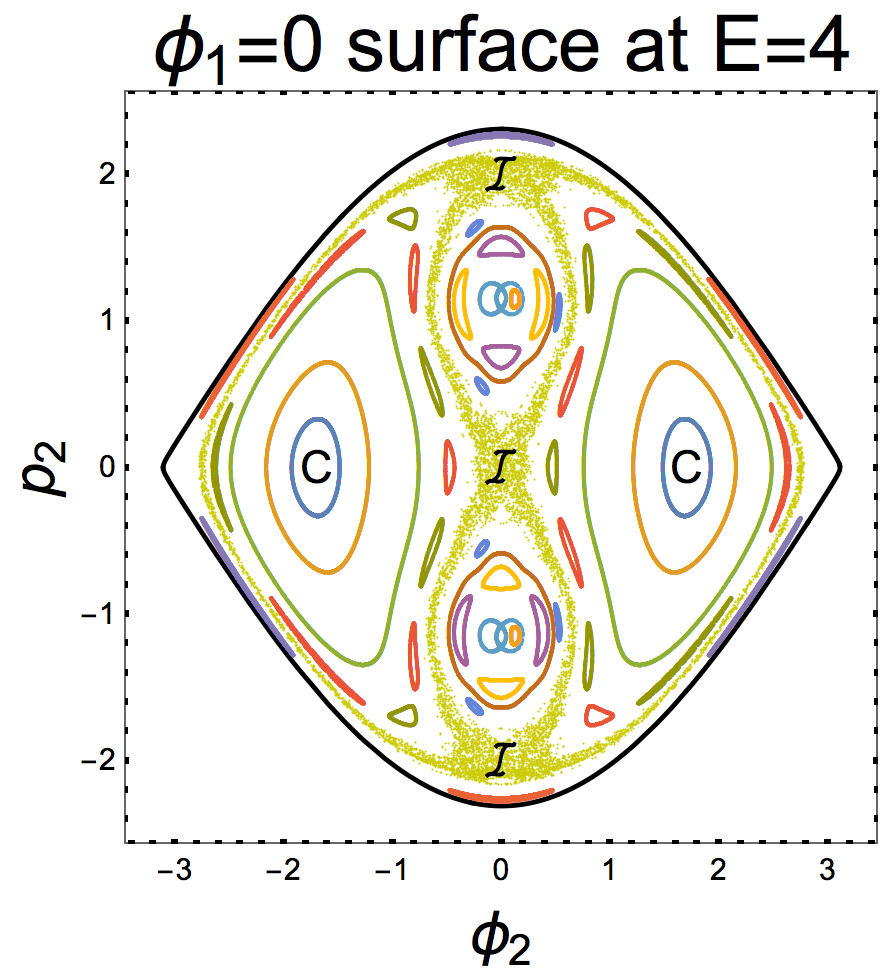}
		\caption{}
		\label{f:psec-egy=4}
	\end{subfigure}	
\quad
	\begin{subfigure}[t]{5.7cm}
		\includegraphics[width=5.7cm]{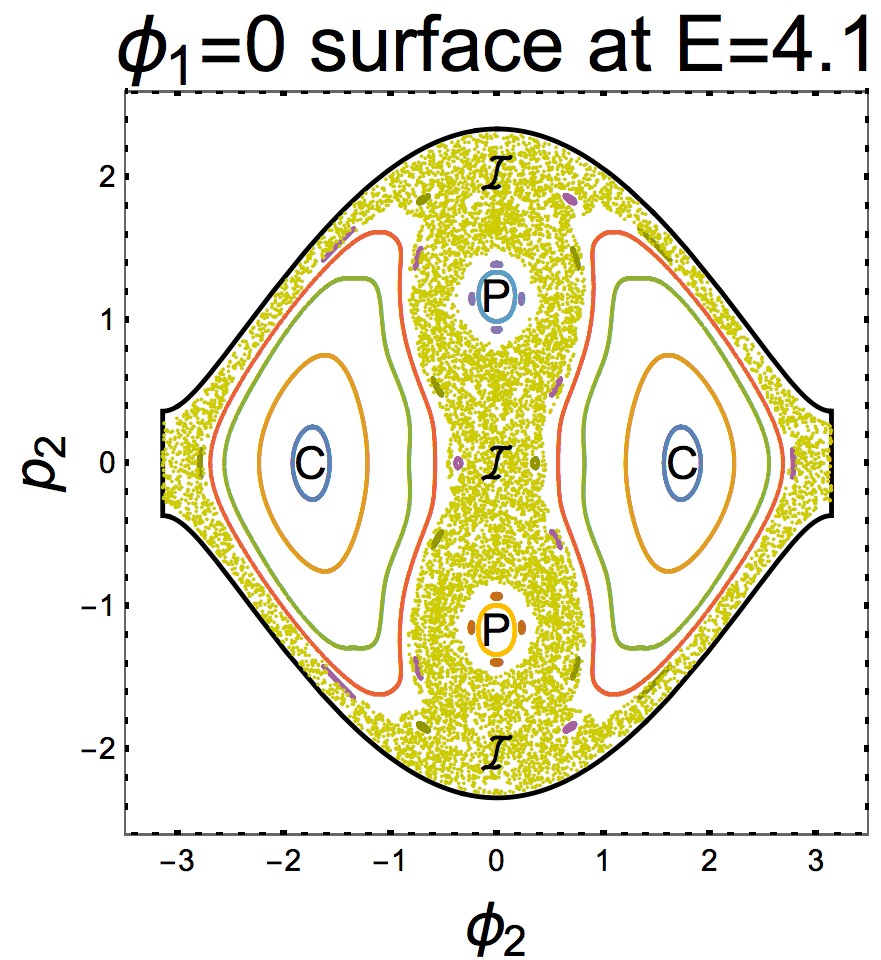}
		\caption{}
		\label{f:psec-egy=4.1}
	\end{subfigure}	
	\caption{  Several Poincar\'e sections on the `$\vf_1 = 0$' surface in the vicinity of $E = 4$ where the chaotic region (shaded, yellow  online) makes its first significant appearance. Distinct sections have different colors online. On each surface, one sees breaking of both up-down and left-right symmetries. Aside from a couple of exceptions on the $E=4$ surface, the set of ICs is left-right and up-down symmetric. The boundary of the Hill region on the `$\vf_1 = 0$' Poincar\'e surface is the $\vf_1 = 0$ pendulum solution. It becomes disconnected for $E > 4$ owing to the bifurcation of the librational pendula into clockwise and counterclockwise rotational pendula.}
	\label{f:psec-egy-near-4}
\end{figure*}

We find that for $E \lesssim 4$, all Poincar\'e sections (on the surface `$\vf_1 = 0$') are nearly regular and display left-right $(\vf_2 \to - \vf_2)$ and up-down $(p_2 \to - p_2)$ symmetries  (see Fig. \ref{f:psec-egy=2-3}). Though there are indications of chaos even at these energies along the periphery of the four stable lobes (e.g., near the unstable isosceles fixed points $\cal I$), chaotic sections occupy a negligible portion of the Hill region. Chaotic sections make their first significant appearance at $E \approx 4$ along the figure-8 shaped separatrix and along the outer periphery of the regular `lobes' that flank it (see Fig. \ref{f:psec-egy-near-4}). This transition to chaos is accompanied by a spontaneous breaking of both the above symmetries. Interestingly, the $\vf_2 \to - \vf_2$ symmetry (though not $p_2 \to - p_2$) seems to be restored when $E \gtrsim 4.4$. The lack of $p_2 \to - p_2$ symmetry at high energies is not unexpected: rotors at high energies either rotate clockwise or counter-clockwise.

\begin{figure*}	
	\begin{subfigure}[t]{3.5cm}
		\centering
		\includegraphics[width=3.5cm]{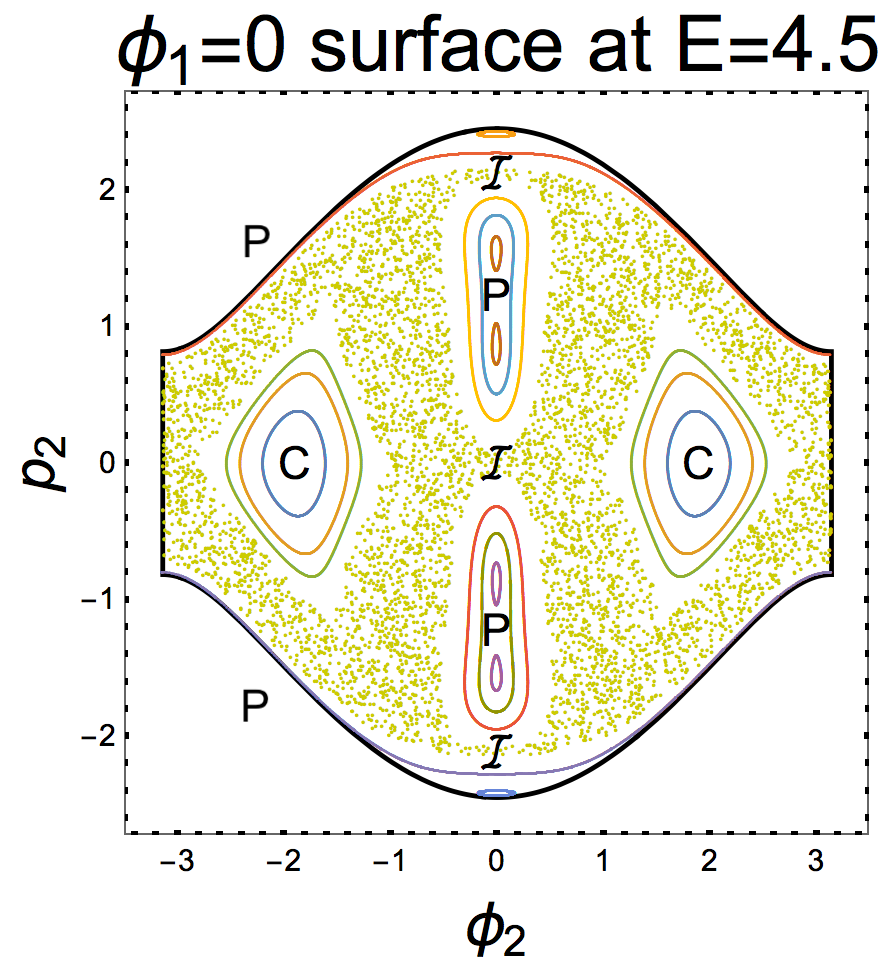}
		\caption{}
		\label{f:psec-egy=4.5}
	\end{subfigure}		
	\begin{subfigure}[t]{3.5cm}
		\centering
		\includegraphics[width=3.5cm]{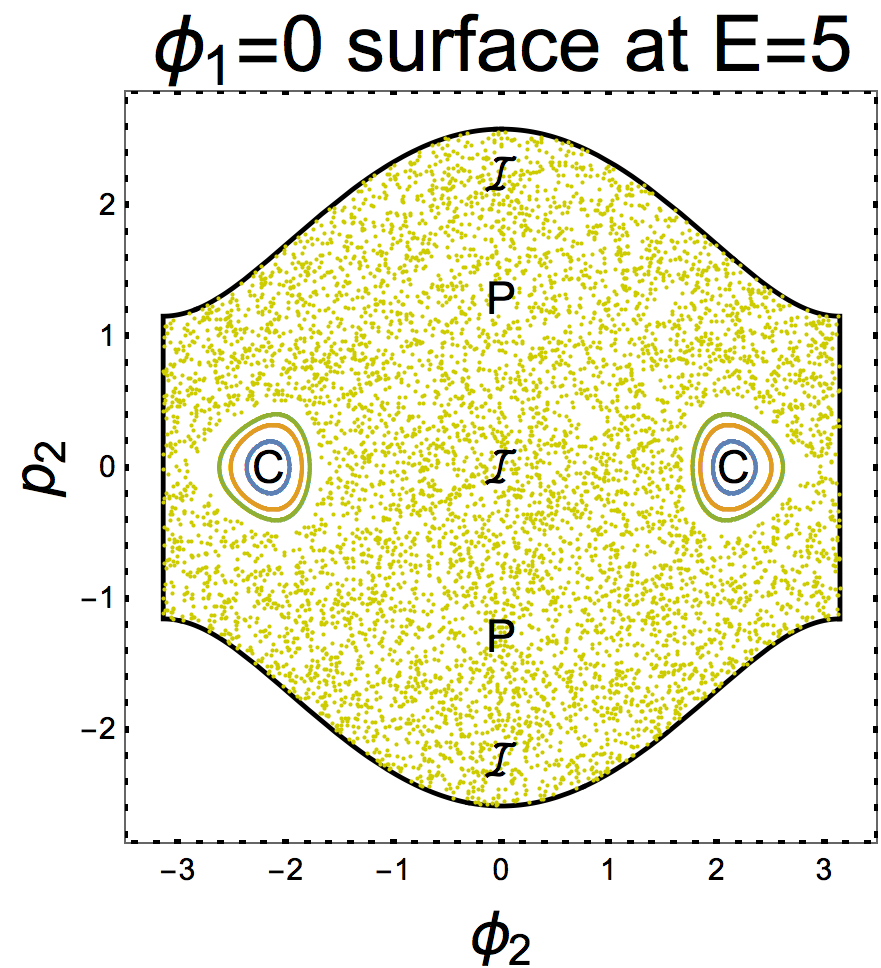}
		\caption{}
		\label{f:psec-egy=5}
	\end{subfigure}	
	\begin{subfigure}[t]{3.5cm}
		\centering
		\includegraphics[width=3.5cm]{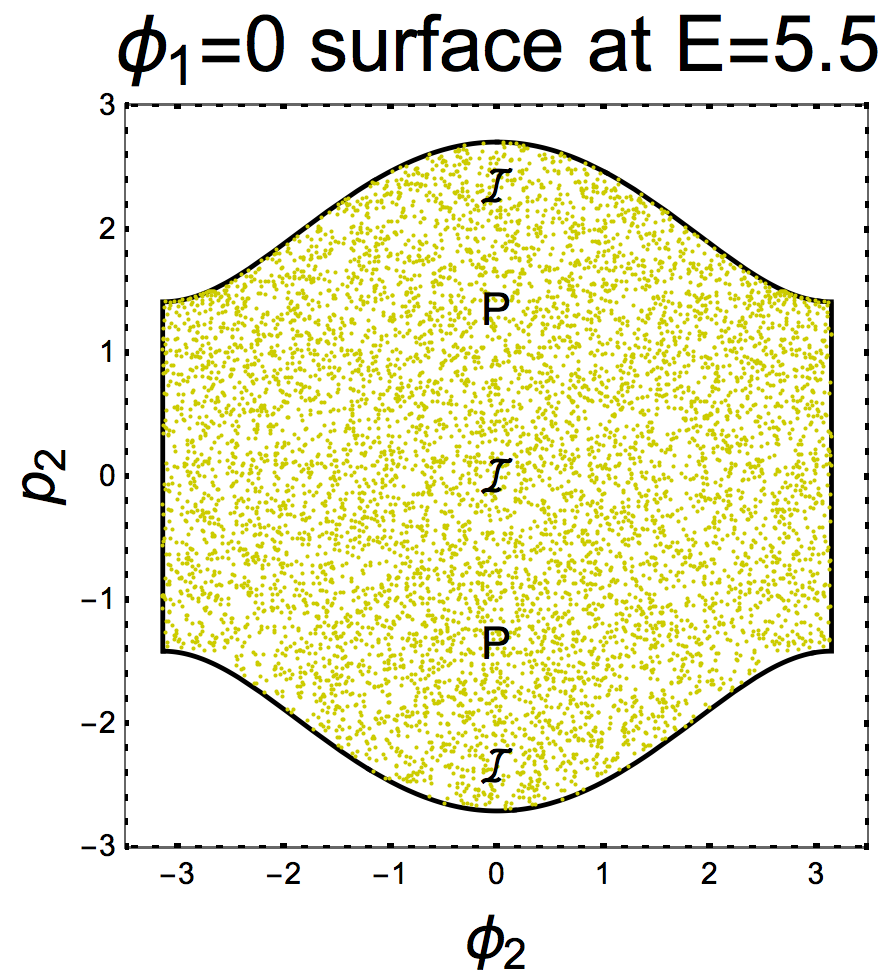}
		\caption{}
		\label{f:psec-egy=5.5}
	\end{subfigure}	
	\begin{subfigure}[t]{3.5cm}
		\centering
		\includegraphics[width=3.5cm]{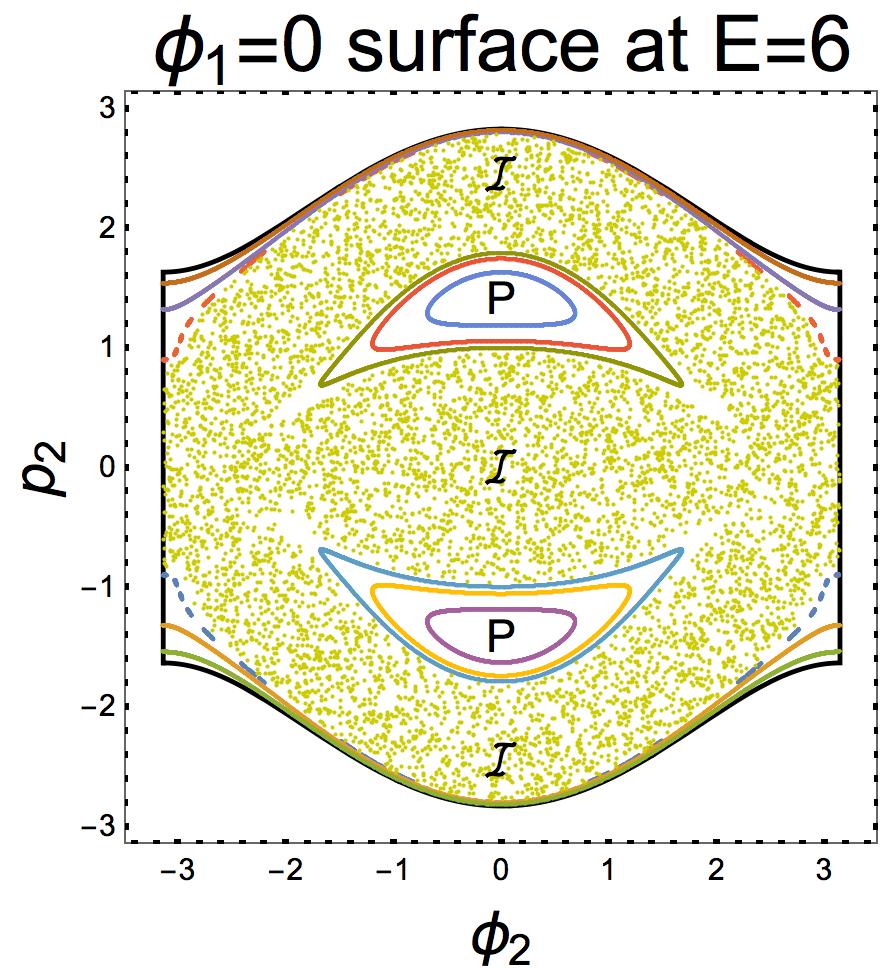}
		\caption{}
		\label{f:psec-egy=6}
	\end{subfigure}
	\begin{subfigure}[t]{3.5cm}
		\centering
		\includegraphics[width=3.5cm]{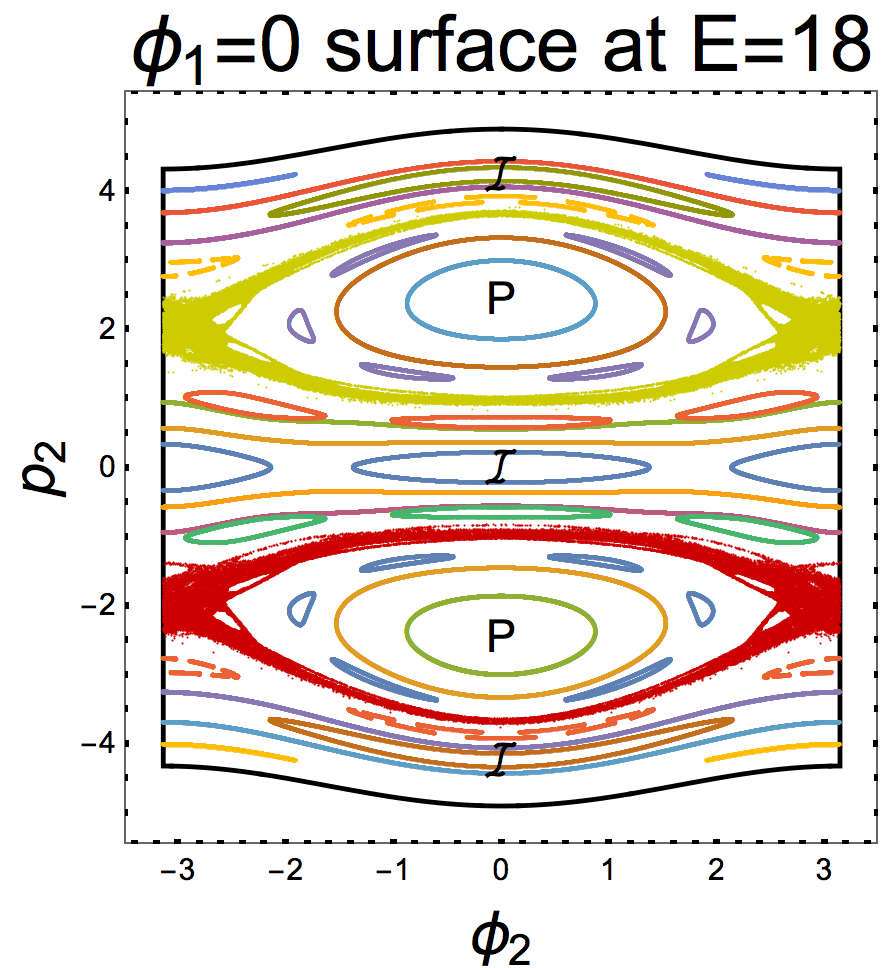}
		\caption{}
		\label{f:psec-egy=18}
	\end{subfigure}	
	\caption{  The up-down symmetry remains broken, though the left-right symmetry is restored on Poincar\'e plots at higher energies. The periodic orbits corresponding to points marked C are choreographies for $E \lesssim 5.33$.}
	\label{f:psec-egy=4.5-to-18}
\end{figure*}

At moderate energies $E \gtrsim 4$, we observe that all chaotic sections (irrespective of the ICs) occupy essentially the same region, as typified by the examples in Fig. \ref{f:psec-egy=4.5-to-18}. At somewhat higher energies (e.g. $E = 14$), we find chaotic sections that fill up both the entire chaotic region and portions thereof when trajectories are evolved up to $t = 10^5$. At yet higher energies (e.g. $E = 18$, Fig. \ref{f:psec-egy=18}), there is no single chaotic section that occupies the entire chaotic region as the $p_2 \to - p_2$ symmetry is broken. 

\subsubsection{Fraction of chaos and global chaos} 

For a range of energies beyond $4$, we find that the area of the chaotic region increases with $E$ (see Fig.~\ref{f:psec-egy-near-4} and \ref{f:psec-egy=4.5-to-18}). At $E \approx 5.5$, the chaotic region coincides with the energetically allowed portion of the Poincar\'e surface (see Fig.~\ref{f:psec-egy=5.5}). Beyond this energy, chaotic sections are supported on increasingly narrow bands (see Fig. \ref{f:psec-egy=18}). This progression towards regular sections is expected since the system acquires an additional conserved quantity in the limit $E \to \infty$. To quantify these observations, we find the `fraction of chaos' $f$ by exploiting the feature that the density of points in chaotic sections is roughly uniform for {\it all energies} on the `$\vf_1 = 0$' surface (this is not true for most other Poincar\'e surfaces). Thus $f$ is estimated by calculating the fraction of the area of the Hill region covered by chaotic sections (see Appendix \ref{a:estimate-f} and Fig. \ref{f:chaos-vs-egy}).

\begin{figure}
	\includegraphics[width=8cm]{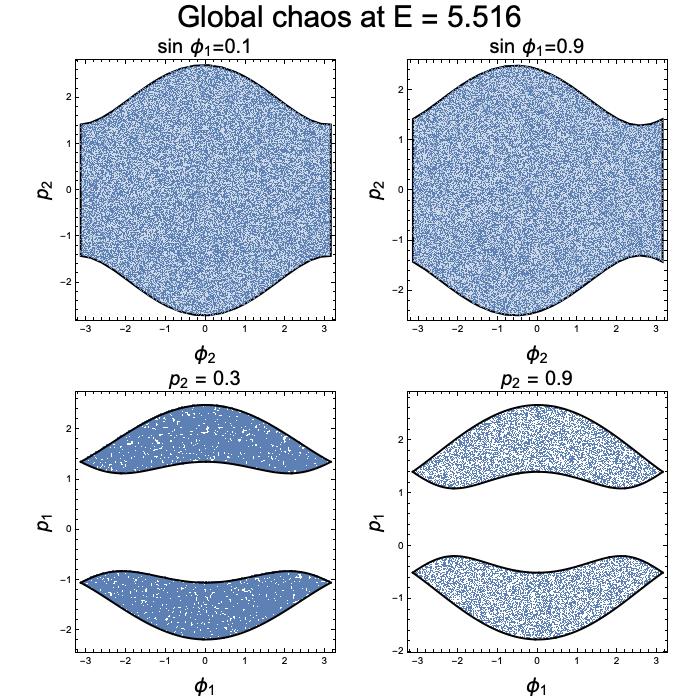}
	\caption{\label{f:global-chaos} Various Poincar\'e surfaces showing global chaos at $E = 5.516$.}
\end{figure}

The near absence of chaos is reflected in $f$ approximately vanishing for $E \lesssim 3.8$. There is a rather sharp transition to chaos around $E \approx 4$ ($f \approx 4\%$, $20\%$ and $40\%$ at $E = 3.85$, $4$ and $4.1$; see lower inset of Fig. \ref{f:chaos-vs-egy}). This is a bit unexpected from the viewpoint of KAM theory and might encode a novel mechanism by which KAM tori break down in this system. Thereafter, $f$ rapidly rises and reaches the maximal value $f \approx 1$ at $E \approx 5.33$. As illustrated in the upper inset of Fig. \ref{f:chaos-vs-egy}, this `fully chaotic' phase persists up to $E \approx 5.6$. Interestingly, we find that for this range of energies, $f \approx 1$ on a variety of Poincar\'e surfaces examined (see Fig. \ref{f:global-chaos}), so that this may be regarded as a phase of `global chaos'. Furthermore, the density of points is uniform on all Poincar\'e surfaces in this phase of global chaos indicating some sort of ergodicity. Additionally, the pendula and breathers are unstable in this phase (see \S \ref{s:reduction-one-dof}) and it would be interesting to know whether this is the case with all periodic solutions. Remarkably, the cessation of the band of global chaos happens to coincide with the energy $E_1^\err \approx 5.6$ above which pendulum solutions are always stable (see Fig. \ref{f:monodromy-evals}). Beyond $E \approx 5.6$, $f$ decreases gradually to zero as $E \to \infty$. Interestingly, the sharp transition to chaos at $E \approx 4$ is also reflected in the JM curvature of \S \ref{s:JM-approach} going from being positive for $E < 4$ to admitting both signs for $E>4$. It is noteworthy that the stable to unstable transition energies in pendula also accumulate from both sides at $E=4$ (see Fig. \ref{f:monodromy-evals}).

\begin{figure}	
	\includegraphics[width=8.2cm]{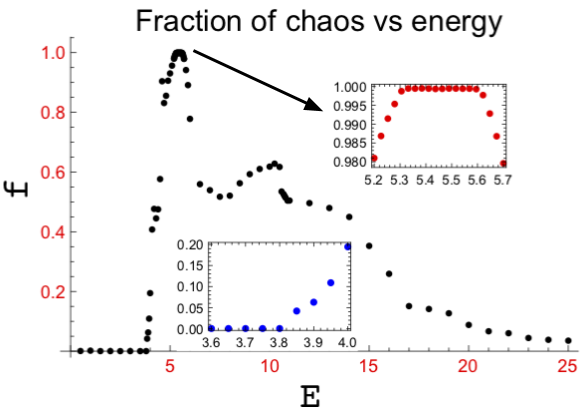}	
	\caption{\label{f:chaos-vs-egy}  Energy dependence of the area of the chaotic region on the `$\vf_1 = 0$' Poincar\'e surface as a fraction of the area of the Hill region.}
\end{figure}

\subsection{Periodic solutions on the Poincar\'e surface `$\vf_1 = 0$'}
\label{s:periodic-sol-on-Psec}

Here, we identify the points on the Poincar\'e surface corresponding to the periodic pendulum and isosceles solutions. Remarkably, careful examination of the Poincar\'e sections also leads us to a new family of periodic `choreography' solutions which are defined and discussed further in \S \ref{s:choreographies}.

\subsubsection{Pendula}

 The $\vf_1 = 0$ pendulum solutions are everywhere tangent to the Poincar\'e surface `$\vf_1 = 0$' and interestingly constitute the `Hill' energy boundary (see Fig. \ref{f:psec-egy=2-3}-\ref{f:psec-egy=4.5-to-18}). [Nb. This connection between pendulum solutions and the Hill boundary is special to the surfaces `$\vf_1 = 0$' and `$\vf_2 = 0$'.] By contrast, the other two classes of pendulum trajectories ($\vf_2 = 0$ and $\vf_1 + \vf_2 = 0$) are transversal to this surface, meeting it at the pendulum points P($0, \pm \sqrt{E/3}$) halfway to the boundary from the origin. These are period-2 and period-1 fixed points for  librational and rotational solutions respectively. Examination of the Poincar\'e sections indicates that pendulum solutions must be stable for $E \lesssim 3.9$ and $E \gtrsim 5.6$ leaving open the question of their stability at intermediate energies. As discussed in \S \ref{s:pendulum-soln}, the pendulua go from being stable to unstable infinitely often as $E \to 4^\pm$. Additionally, by considering initial conditions near the pendulum points, we find that the pendulum solutions lie within the large chaotic section only between $E \approx 4.6$ and the cessation of global chaos at $E \approx 5.6$.

\subsubsection{Breathers}

 Unlike pendula, all isosceles periodic orbits intersect the `$\vf_1 = 0$' surface transversally at points on the vertical axis. Indeed, the breathers defined by $\vf_1 = \vf_2$ and $\vf_2 + 2\vf_1 = 0$ intersect the surface at the isosceles points ${\cal I}(0, \pm \sqrt{E})$ which form a pair of period-2 fixed points for $E < 4.5$ and become period-1 in the rotational regime (see Fig. \ref{f:psec-egy=2-3}-\ref{f:psec-egy=4.5-to-18}). The breathers defined by $\vf_1 + 2\vf_2 = 0$ intersect the surface at the period-1 fixed point at the origin. In agreement with the conclusions of \S \ref{s:stability-of-breathers}, the Poincar\'e sections show that all three isosceles points are unstable at low energies, lie in the large chaotic section for $3.9 \lesssim E \lesssim 8.97$ and are stable at higher energies.

\subsubsection{A new family of periodic solutions}

 The period-2 fixed points C at the centers of the right and left lobes on the Poincar\'e surfaces of Fig. \ref{f:psec-egy=2-3} and \ref{f:psec-egy-near-4} correspond to a new family of periodic solutions. Evidently, they go from being stable to unstable as the energy crosses $E \approx 5.33$. We argue in \S \ref{s:choreographies} that they are choreographies for $E \lesssim 5.33$.

\section{Choreographies} 
\label{s:choreographies}

\begin{figure}	
	\centering
	\begin{subfigure}[t]{4.3cm}
		\centering
		\includegraphics[width=4.3cm]{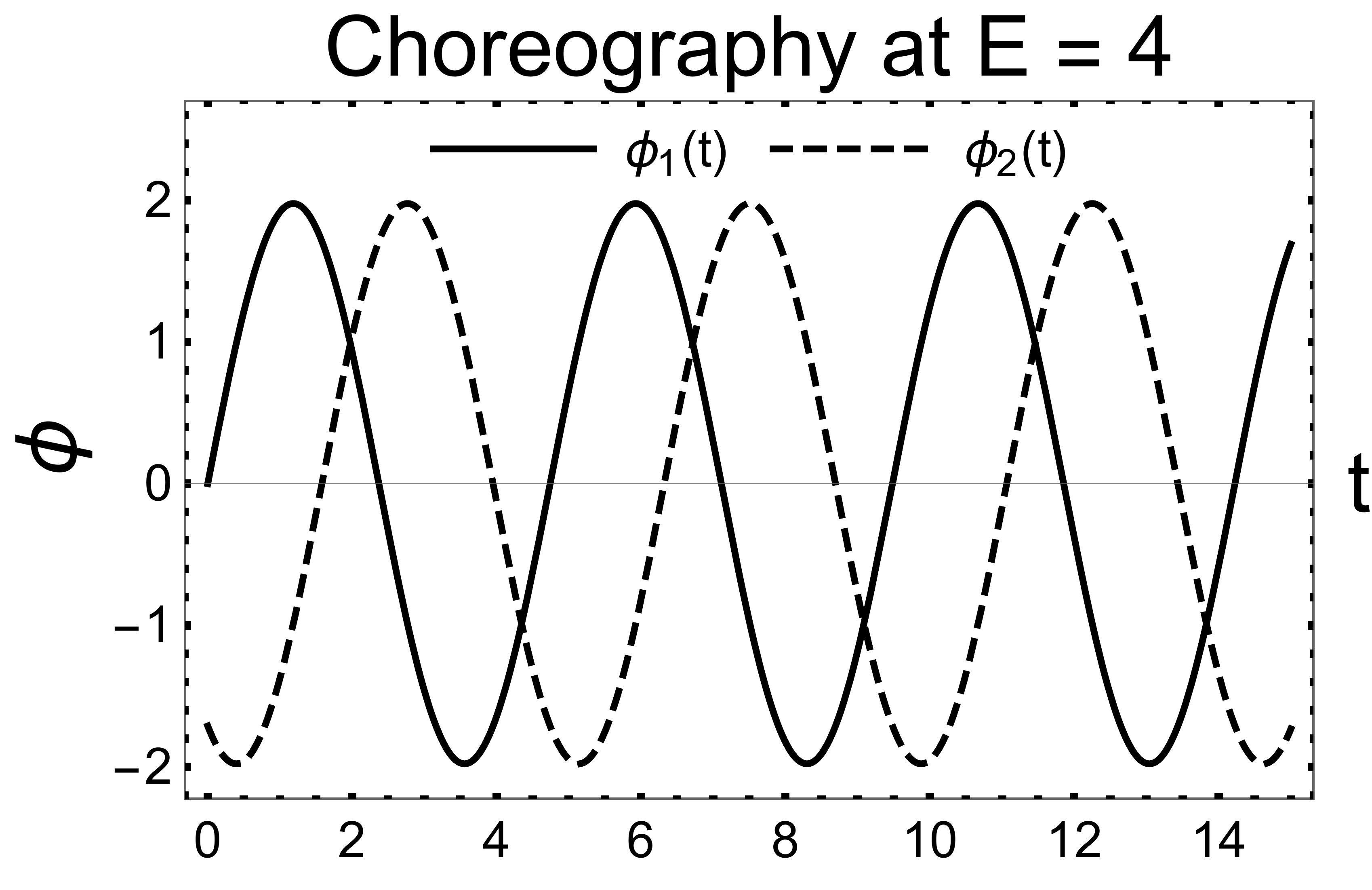}
		\caption{}
		\label{f:choreo-phi1-phi2-plot}
	\end{subfigure}
\qquad
	\begin{subfigure}[t]{3.5cm}
		\centering
		\includegraphics[width=3.5cm]{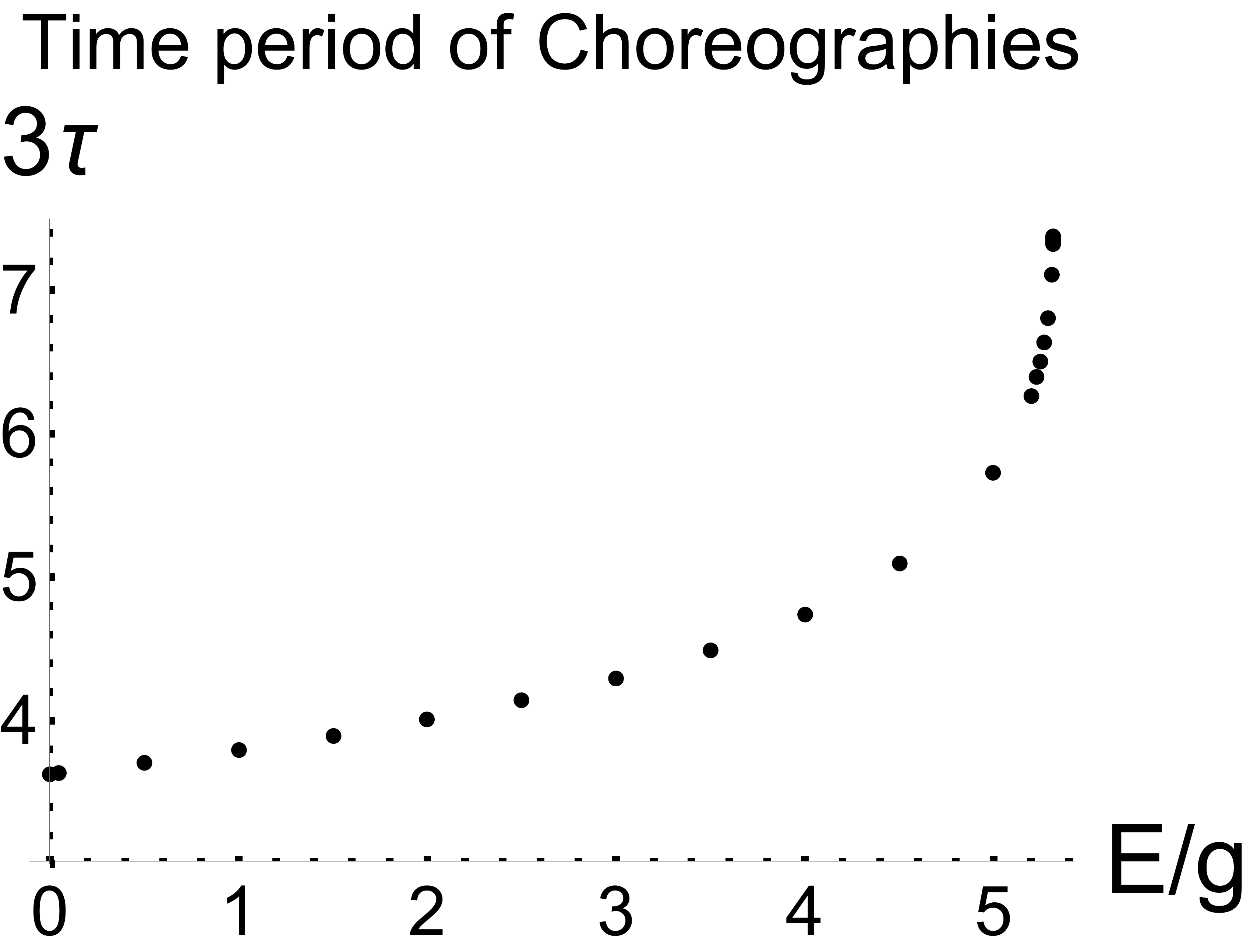}
		\caption{}
		\label{f:choreo-time-vs-egy}
	\end{subfigure}
	\caption{  (a) A non-rotating choreography at $E = 4g$ showing that the time lag between $\vf_1$ and $\vf_2$ is one-third the period. (b) The time period $3\tau$ of non-rotating choreographies as a function of energy indicating divergence at $E \approx 5.33g$.}
	\label{f:choreo-time-vs-egy-phi1-phi2-plot}
\end{figure}

Choreographies are an interesting class of periodic solutions of the $n$-body problem where all particles follow the same closed curve equally separated in time \cite{montgomery-choreographies}. The Lagrange equilateral solution where three equal masses move on a common circle and the stable zero-angular momentum figure-8 solution discovered by C. Moore \cite{crismoore} (see also Ref.\cite{chenciner-montgomery}) are perhaps the simplest examples of choreographies in the equal mass gravitational 3 body problem. Here, we consider choreographies in the 3 rotor problem where the angles $\tht_i(t)$ of the three rotors may be expressed in terms of a single $3\tau$-periodic function, say $\tht_1(t)$:
	\beq
	\label{e:defn-choreography}
	\tht_2(t) = \tht_1(t+\tau) \;\; 
	\text{and} 
	\;\; \tht_3(t) = \tht_1(t + 2\tau).
	\eeq
This implies that the CM and relative coordinates $\vf_0$, $\vf_1(t)$ and $\vf_2(t) = \vf_1(t+\tau)$ must be $3\tau$ periodic (see Fig. \ref{f:choreo-phi1-phi2-plot}) and satisfy the delay algebraic equation
	\beqs
	\vf_1(t) + \vf_1(t+\tau) + \vf_1(t+2\tau) &=& \cr 
	\tht_1 - \tht_2 + \tht_2 - \tht_3 + \tht_3 - \tht_1 &\equiv& 0 \mod 2\pi.
	\label{e:choreography-delay-algebraic}
	\eeqs
The EOM (\ref{e:3rotors-EOM-ph1ph2}) become $3 m r^2 \ddot \vf_0 = 0$ and the pair of delay differential equations
	\beqs
	mr^2 \ddot \vf_1(t) &=& - g\big[ 2 \sin \vf_1(t) - \sin \vf_1(t+\tau) \cr
	&& + \sin (\vf_1(t) + \vf_1(t+\tau)) \big] \quad \text{and} \cr 
	mr^2 \ddot \vf_2(t) &=& mr^2 \ddot \vf_1(t+\tau) = -g \big[ 2 \sin \vf_1(t+\tau) \cr
	&&- \sin \vf_1(t) + \sin (\vf_1(t) + \vf_1(t+\tau)) \big].
	\label{e:choreography-phi-1-2}
	\eeqs
In fact, the second equation in (\ref{e:choreography-phi-1-2}) follows from the first by use of the delay algebraic equation (\ref{e:choreography-delay-algebraic}). Moreover, using the definition of $\vf_0$, the constant angular velocity of the CM
	\beqs
	\dot \vf_0 &=& \ov{\tau} \left[\vf_0(t+\tau) - \vf_0(t)\right] \cr
	&=& - \fr{1}{3\tau} \left[ \vf_1(t) + \vf_1(t+\tau) + \vf_1(t+2\tau)\right]. 
	\label{e:choreography-CM-angular-velocity}
	\eeqs
It is verified that any $3\tau$ periodic triple $\vf_{0,1,2}$ satisfying (\ref{e:choreography-delay-algebraic}), (\ref{e:choreography-phi-1-2}) and (\ref{e:choreography-CM-angular-velocity}) leads to a choreography of the 3-rotor system. Thus, to discover a choreography we only need to find a $3\tau$-periodic function $\vf_1$ satisfying (\ref{e:choreography-delay-algebraic}) and the first of the delay differential equations (\ref{e:choreography-phi-1-2}) with the period $3 \tau$ self-consistently determined. Now, it is easy to show that choreographies cannot exist at asymptotically high (relative) energies. In fact, at high energies, we may ignore the interaction terms ($\propto g$) in (\ref{e:choreography-phi-1-2}) to get $\vf_1(t) \approx \om t + \vf_1(0)$ for $|\om| \gg 1$. However, this is inconsistent with (\ref{e:choreography-delay-algebraic}) which requires $3 \om t \equiv 0 \mod 2\pi$ at all times. On the other hand, as discussed below, we {\it do} find examples of choreographies at low and moderate relative energies.

\subsection{Examples of choreographies}

 Uniformly rotating (at angular speed $\Om$) versions of the static solutions G and T (but not D) (see \S \ref{s:rotating-static-sol} and Fig.~\ref{f:static-solutions-3rotors}) provide the simplest examples of choreographies with $\tht_1(t) = \Om t$ and $\tau = 2\pi/\Om$ for G and $\tau = 2\pi/3\Om$ for T where $\Om$ is arbitrary. In the case of G, though all particles coincide, they may also be regarded as separated by $\tau$. The energies (\ref{e:egy-3rotors-phi1-phi2-coords}) of these two families of choreographies come from the uniform CM motion and a constant relative energy:
	\beq
	E^\text{(G)}_{\rm tot} = \fr{3}{2} m r^2 \Om^2 \quad \text{and} \quad E^\text{(T)}_{\rm tot} = \fr{3}{2} m r^2 \Om^2 + \fr{9g}{2}.
	\eeq
These two families of choreographies have the scaling property: if $\tht(t)$ with period $3\tau$ describes a choreography in the sense of (\ref{e:defn-choreography}), then $\tht(a t)$ with period $|3\tau/a|$ also describes a choreography for any real $a$. It turns out that the above two are the only such `scaling' families of choreographies. To see this, we note that both $\tht(t)$ and $\tht(at)$ must satisfy the delay differential equation
	\beqs
	&\ddot \tht(t+\tau) - \ddot \tht(t) = \fr{-g}{mr^2} \bigg[ 2 \sin (\tht(t+\tau) - \tht(t))  \cr
	& - \sin (\tht(t) - \tht(t-\tau)) + \sin (\tht(t+\tau) - \tht(t-\tau)) \bigg]
	\eeqs
implying that either $a^2 = 1$ or $\ddot \tht(t + \tau) = \ddot \tht(t)$. However, the latter implies that $\dot \tht(t+\tau) - \dot \tht(t) = -\dot \vf_1(t)$ is a constant which must vanish for the delay algebraic equation (\ref{e:choreography-delay-algebraic}) to be satisfied. Consequently, $\dot \vf_2$ must also vanish implying that the choreography is a uniformly rotating version of G or T.

\subsection{Non-rotating choreographies}

 Remarkably, we have found another 1-parameter family of choreographies (e.g., Fig. \ref{f:choreo-phi1-phi2-plot}) that start out as small oscillations around G. At low energies, they have a period $3 \tau = 2\pi/\om_0$ and reduce to 
	\beq
	\vf_1(t) \approx \sqrt{\fr{2E}{3g}} \sin (\om_0 (t - t_0)) \quad \text{for} \quad E \ll g
	\label{e:choreography-low-egy}
	\eeq
where $\om_0 = \sqrt{3 g/mr^2}$. It is easily verified that (\ref{e:choreography-delay-algebraic}) is identically satisfied while  (\ref{e:choreography-phi-1-2}) is satisfied for $E \ll g$. Moreover, using (\ref{e:choreography-CM-angular-velocity}), we find that the angular speed $\dot \vf_0$ of the CM must vanish for (\ref{e:choreography-low-egy}) so that the energy is purely from the relative motion. The phase trajectory corresponding to (\ref{e:choreography-low-egy}) intersects the $\vf_1 = 0$ Poincar\'e surface at the pair of period-2 fixed points C$(\pm \sqrt{E/2g},0)$ which lie at the centers of the left and right stable `lobes' pictured in Fig. \ref{f:psec-egy=2-3} at $E = 2g$ and $3g$. 

More generally, we numerically find that when the ICs are chosen at the stable fixed points at the centers of these lobes, the trajectories are a one-parameter family of choreographies $\vf_1(t;E)$ varying continuously with $E$ up to $E \approx 5.33$. It can be argued that these choreographies are non-rotating (involve no CM motion). Indeed, from (\ref{e:choreography-CM-angular-velocity}) and (\ref{e:choreography-delay-algebraic}), we must have  $3 \tau \dot \vf_0 \equiv 0 \mod 2\pi$, implying that $\dot\vf_0$ cannot jump discontinuously. Since, $3 \tau \dot \vf_0 = 0$ as $E \to 0$ (\ref{e:choreography-low-egy}), it must remain zero when $E$ is continuously increased from $0$ to $5.33$. Though we do not study their stability here by the monodromy approach, the Poincar\'e sections (see Fig.~\ref{f:psec-egy=2-3} and \ref{f:psec-egy-near-4}) indicate that they are stable. As shown in Fig. \ref{f:choreo-time-vs-egy}, the time period $3 \tau$ grows monotonically with $E$ and appears to diverge at $E \approx 5.33$, which coincides with the beginning of the band of `global chaos' (see \S \ref{s:poincare-section}). For $E \gtrsim 5.33$, the period-2 choreography points C on the `$\vf_1 = 0$' Poincar\'e surface become unstable   and lie in a chaotic region (see Fig. \ref{f:psec-egy=4.5-to-18}), preventing us from finding such a choreography, if it exists, using the above numerical technique. As argued before, choreographies are forbidden at very high energies. For instance, on the `$\vf_1 = 0$' Poincar\'e surface at $E = 18$ (see Fig. \ref{f:psec-egy=18}), the analogues of the C points correspond to unstable periodic orbits which are {\it not} choreographies. In fact, we conjecture that this family of periodic solutions ceases to be a choreography beyond $E \approx 5.33$.

\section{Discussion}
\label{s:discussion}

In this paper, we have studied the classical three rotor problem and found novel signatures of its transition to chaos as well as a phase of global chaos. We also discovered `pendulum' and `isosceles-breather' periodic solutions as well as choreographies and discussed their stability properties. \S \ref{s:introduction} contains a concise summary of our results. Here, we discuss some open questions arising from our work.

The classical 3 rotor problem and planar restricted 3 body problem are similar in the sense that both have essentially two degrees of freedom and only one known conserved quantity. In the latter, Bruns and Poincar\'e \cite{whittaker} proved the non-existence of additional conserved quantities of certain types (analytic in small mass ratios and orbital elements). It would be reassuring to obtain a similar result for the 3 rotor problem. Analogously, the extension to our system, of Ziglin's and Melnikov's arguments for non-integrability is also of interest \cite{ziglin,melnikov}.

While we found the trace of the monodromy for periodic `pendulum' solutions numerically, it would be interesting to prove the accumulation of stable to unstable phase transitions at $E = 4g$  as in Ref. \cite{churchill} and establish its asymptotic periodicity on a log scale, for instance by finding an analytical expression for the stability index as Yoshida \cite{yoshida84} does in the 2d anharmonic oscillator of Eq. (\ref{e:anharmonic-oscillator-yoshida}). This accumulation at the threshold for bound librational trajectories with diverging time periods and the periodicity on a log scale is reminiscent of the quantum energy spectrum of Efimov trimers that accumulate via a geometric sequence at the two-body bound state threshold with diverging S-wave scattering length \cite{efimov-original}. It would also be interesting to explore a possible connection between this accumulation of transitions and the accumulation of homoclinic points at a hyperbolic fixed point in a chaotic system. The nature of bifurcations \cite{brack-mehta-tanaka} and local scaling properties \cite{LSS} at these transitions are also of interest. In another direction, one would like to understand if there is any connection between the accumulation of transition energies and the change in topology of the Hill region $(V \leq E)$ of the configuration torus as $E$ crosses the value $4g$ at the three critical points (saddles D) of the Morse function $V$ (see \S \ref{s:topology-hill-region}). One would also like to analyse the onset of widespread chaos in this system using methods such as those of Chirikov\cite{chirikov} and Greene\cite{green}.


We have argued that the 3 rotor system is integrable at $E = 0$ and $\infty$ ($g = \infty, 0$), where  additional conserved quantities emerge. One wonders whether it is `integrable' at any other energy. In other words, is there any non-trivial energy hypersurface in phase space on which all trajectories are periodic or quasi-periodic so that the corresponding Poincar\'e sections are regular? Our estimate of the fraction of chaos on the `$\vf_1 = 0$' Poincar\'e surface strongly suggests that any integrable energy $E_{\rm I}$ is either isolated or $E_{\rm I} \lesssim 3.8g$. However, even for low energies, we expect chaotic sections in the neighborhood of the isosceles points $\cal I$ (see Fig. \ref{f:psec-egy=2-3}). In fact, we conjecture that the 3 rotor problem has no non-trivial integrable energies unlike the 2d anharmonic oscillator\cite{yoshida84}. 

As discussed in \S \ref{s:poincare-section}, Poincar\'e sections suggest a band of global chaos for $5.33g \lesssim E \lesssim 5.6g$. This is of course consistent with the instability of pendulum and breather solutions in this regime. Consequently, it would be interesting to investigate the possible ergodic behavior of three rotors for such energies.

Finally, a deeper understanding of the physical mechanisms underlying the onset of chaos in this system would be desirable, along with an examination of quantum manifestations of the classical chaos, given the connection to modeling chains of coupled Josephson junctions.

\begin{acknowledgments}
We thank K G Arun, M Berry, A Chenciner, S Dattagupta, S R Jain, A Lakshminarayan, R Nityananda, T R Ramadas, M S Santhanam and anonymous reviewers for helpful comments and references. This work was supported in part by the Infosys Foundation, J N Tata trust and grants (MTR/2018/000734, CRG/2018/002040) from the Science and Engineering Research Board, Govt. of India.
\end{acknowledgments}

\appendix

\section{Quantum $N$-rotor problem from  XY model} 
\label{a:n-rotor-from-xy}

The quantum $N$-rotor problem may be related to the 2d XY model of classical statistical mechanics which displays the celebrated Kosterlitz-Thouless topological phase transition \cite{sachdev}.
The dynamical variables of the XY model are 2d unit-vector spins ${\bf S}_\al$ (or phases $e^{i \tht_\al}$) at each site $\al$ of an $N \times M$ rectangular lattice with horizontal and vertical spacings $a$ and $b$ and nearest neighbor  ferromagnetic interaction energies $- J \: {\bf S}_\al \cdot {\bf S}_\beta = - J \cos (\tht_\al - \tht_\beta)$ with $J > 0$. One often considers $a=b$ and assumes that $\tht$ varies gradually so that in the continuum limit $a \to 0$ and $N, M \to \infty$ holding $aN$ and $aM$ fixed, the Hamiltonian becomes $H = \fr{J}{2} \int |\grad \tht|^2 \: d^2{\bf r}$. This defines the 1+1 dimensional $O(2)$ principal chiral model.

Here, we approximately reformulate the XY model as an interacting quantum $N$-rotor problem by taking a partial continuum limit in the vertical direction followed by a Wick rotation. The resulting quantum system has been used to model a 1d array of coupled Josephson junctions and is known to be related to the XY model in a Villain approximation \cite{sondhi-girvin,wallin-1994}. With $i$ and $j$ labelling the columns and rows of the lattice, the XY model Hamiltonian is 
	\beq
	H = - J \sum_{i,j} \left[ \cos(\tht_{i,j+1}-\tht_{i,j}) + \cos(\tht_{i+1,j}-\tht_{i,j}) \right]  
	\label{e:Hamiltonian-XY}
	\eeq
with $J > 0$. In the first term, the sum is over $1\leq i \leq N$ and $1\leq j \leq M-1$ while for the second term, we have $1 \leq i \leq N-1$ and $1 \leq j \leq M$. We will impose periodic boundary conditions (BCs) in the horizontal but not in the vertical direction (open BCs are also of interest). We will take a continuum limit in two steps. We first make the spacing between rows small by introducing a continuous vertical coordinate $\tau$ in place of $j$ such that $\tau(j+1) - \tau(j) = \del \tau = b$. Next, we approximate $\cos(\tht_{i,j+1}-\tht_{i,j})$ by
	\beqs
	\cos(\tht_{i}(\tau+\del \tau)-\tht_i(\tau)) &\approx& 1-\half (\tht_i(\tau + \del \tau)-\tht_i(\tau))^2 \cr
	&\approx& 1- \half \tht_i'(\tau)^2 \, b \, d\tau.
	\eeqs 
Here, we have chosen to write $(\del\tau)^2$ as $b \: d \tau$ in anticipation of taking $b \to 0$ in the second step. 
Within this approximation, the Hamiltonian (\ref{e:Hamiltonian-XY}) up to an additive constant becomes 
	\beq
	H = J \sum_i \int \left\{ \fr{b}{2} \tht_i'(\tau)^2 - \ov{b} \cos\left[\tht_{i+1}(\tau)-\tht_{i}(\tau) \right] \right\} \: d\tau
	\eeq
using the prescription $\sum_j b f(\tau_j) \to \int f(\tau) d\tau$. The resulting partition function
	\beq
	Z = \int \prod_{k=1}^N D[\tht_k] \exp\left[-\beta H \right]
	\eeq
after a Wick rotation $\tau = i c t$, may be written as	
	\beqs
	Z &=& \int D[\tht] e^{i S/\hbar} \quad \text{where} \cr
	\fr{S}{\hbar} &=& \beta J c \sum_i \int dt \left[ \fr{b}{2 c^2} \dot \tht_i(t)^2 + \ov{b} \cos\left[\tht_{i+1}(t)-\tht_{i}(t) \right] \right].
	\label{e:partition-fn-cnt-vertical-discrete-horizontal}
	\eeqs
We introduced a parameter $c > 0$ with dimensions of speed so that $t$ has dimensions of time. We may take a second continuum limit, this time in the horizontal direction by replacing $\sum_i$ by $\int \fr{dx}{a}$ by taking $a \to 0$ and $N \to \infty$ while holding $aN$ and $a/b$ fixed to get
	\beqs
	\fr{S}{\hbar} &\approx& \beta J c \int \fr{dx}{a}  \int dt \left\{ \fr{b}{2c^2} \left( \fr{\pdr \tht}{\pdr t} \right)^2 + \ov{b} \cos\left( a \fr{\pdr \tht}{\pdr x} \right) \right\} \cr
	&\approx& \half \beta J c \int dx \; dt \left\{ \fr{b}{a}\ov{c^2} \dot \theta^2 - \fr{a}{b} \tht'^2 \right\}.
	\eeqs
The path integral $\int D[\tht] e^{i S/\hbar}$ is what we would have obtained if we had taken the conventional continuum limit ($a,b \to 0$) of the XY model partition function and then performed a Wick rotation. Our two-step continuum limit has allowed us to approximately identify the quantum $N$-rotor problem (\ref{e:partition-fn-cnt-vertical-discrete-horizontal}) where $b$ has not yet been taken to zero.

For fixed $N, a$ and $b$, the physical interpretation of (\ref{e:partition-fn-cnt-vertical-discrete-horizontal}) is facilitated by letting $Lb/ac \beta$ play the role of $\hbar$ where $L$ is a length that remains finite in the limit $a, b \to 0$. $L$ could be the horizontal linear dimension of the system. This $\hbar$ has dimensions of action and tends to $0$ at low temperatures where quantum fluctuations in the Wick rotated theory should be small. With this identification of $\hbar$, we read off the classical action
	\beq
	\label{e:classical-action-intermediate-problem}
	S[\tht] = \sum_{i} \int  \left\{ \fr{ J L b^2}{2 a c^2} \dot\tht_i^2 + \fr{J L}{a} \cos\left[\tht_i-\tht_{i+1} \right] \right\} \; dt. 
	\eeq
Letting $m = J/c^2$, $r = \sqrt{L b^2/a}$ and $g = J L/a$, the corresponding Hamiltonian (with $\tht_{N+1} \equiv \tht_1$)
	\beq
	H = \sum_{i=1}^N  \left\{ \half { m r^2} \dot\tht_i^2 + g [1 - \cos\left(\tht_i-\tht_{i+1} \right) ] \right\}
	\eeq
describes the equal mass $N$-rotor problem. The rotor angles $\tht_i$ parametrize $N$ circles whose product  is the $N$-torus configuration space. Though the rotors are identical, each is associated to a specific site and thus are distinguishable. In particular, the wavefunction $\psi(\tht_1, \tht_2, \cdots \tht_N)$ need not be symmetric or antisymmetric under exchanges. We may also visualize the motion by identifying all the circles  but allowing the rotors/particles to remember their order from the chain. So particles $i$ and $j$ interact only if $i-j = \pm 1$. In particular, particles with coordinates $\tht_1$ and $\tht_3$ can freely `pass through' each other! Furthermore, on account of the potential, particles $i$ and $i+1$ can also cross without encountering singularities. Finally, we note that the quantum Hamiltonian corresponding to (\ref{e:classical-action-intermediate-problem}),
	\beq
	\hat H 
	= \sum_i - \fr{\hbar^2}{2 m r^2} \fr{\pdr^2}{\pdr\tht_i^2} - g \cos(\tht_i - \tht_{i+1})
	\eeq
has been used to model a 1d array of coupled Josephson junctions\cite{sondhi-girvin} with the capacitive charging and Josephson coupling energies given by $E_C = \hbar^2/m r^2 = L/a \beta^2 J$ and $E_J = g = JL/a$.

\section{Positivity of JM curvature for $0 \leq E \leq 4g$} 
\label{a:positivity-of-JM-curvature}
 
Here, we prove that for $0 \leq E \leq 4g$, the JM curvature $R$ of \S \ref{s:JM-approach} is strictly positive in the Hill region ($E > V$) of the $\vf_1$-$\vf_2$ configuration torus. It is negative outside and approaches $\pm \infty$ on the Hill boundary $E = V$.  It is convenient to work in Jacobi coordinates $\vf_\pm = (\vf_1 \pm \vf_2)/2$ introduced in \S \ref{s:jacobi-coordinates} and define $P = \cos \vf_+$ and $Q = \cos \vf_-$. In these variables,
 	\beqs
	R = \fr{g^2 {\cal N}_E(P,Q)}{mr^2 (E-V)^3}  \quad &\text{where} \quad {\cal N}_E = 5 + 2Q^2 - 6 P Q + 8 P^3 Q \cr 
	&+  \left[\fr{2E}{g} - 3\right] (2P^2 + 2 P Q - 1).
	\label{e:N}
	\eeqs
Since $E-V > 0$ in the Hill region, it suffices to show that ${\cal N}_E \geq 0$ on the whole torus and strictly positive in the Hill region. It turns out that (a) ${\cal N}_E \geq 0$ for $E = 0$ and $4g$ and (b) for $E = 0$, ${\cal N}_E$ vanishes only at the ground state G while for $E = 4g$, it vanishes only at the saddles D, with both G and the Ds lying on the Hill boundary. Since G is distinct from the Ds, linearity of ${\cal N}_E$  then implies that ${\cal N}_E > 0$ on the entire torus for $0 < E < 4g$. It only remains to prove (a) and (b).

To proceed, we regard ${\cal N}_E$ as a function on the $[-1,1]\times[-1,1]$ $PQ$-square. (i) When $E=0$, ${\cal N}_0$ has only one local extremum in the interior of the $PQ$-square at $(0,0)$ where ${\cal N}_0(0,0) = 8$. On the boundaries of the $PQ$-square,
	\beqs
	{\cal N}_0(\pm 1,Q) &=& 2(1 \mp  Q)^2 \geq 0 
	\quad \text{and} \cr
	{\cal N}_0(P, \pm 1) &=& 2(P\mp1)^2 (5 \pm 4 P) \geq 0
	\eeqs
with ${\cal N}_0$ vanishing only at $(1,1)$ and $(-1,-1)$ both of which correspond to G. Thus, ${\cal N}_0 \geq 0$ on the whole torus and vanishes only at G which lies on the Hill boundary. (ii) When $E = 4g$, the local extrema in the interior of the $PQ$-square are at $(0,0)$ and $(\pm 1, \mp 5/3)/\sqrt{3}$ where ${\cal N}_{4g}$ takes the values $0$ and $40/27$. On the boundaries of the $PQ$-square,
	\beqs
	{\cal N}_{4g}(\pm 1,Q) &=& 2 (1 \pm Q) (5 \pm Q) \geq 0
	\quad \text{and} \cr 
	{\cal N}_{4g}(P,\pm 1) &=& 2 (1 \pm P)(1 \pm P + 4 P^2) \geq 0
	\eeqs
with ${\cal N}_{4g}$ vanishing only at $(1, -1)$ and $(-1, 1)$. Hence, for $E = 4g$,  ${\cal N}_{4g} \geq 0$ on the whole torus and vanishes only at the three saddle points (Ds) all of which lie on the Hill boundary.

\section{Measuring area of chaotic region on `$\vf_1 = 0$' Poincar\'e surface}
\label{a:estimate-f}

\begin{figure}
	\includegraphics[width=8cm]{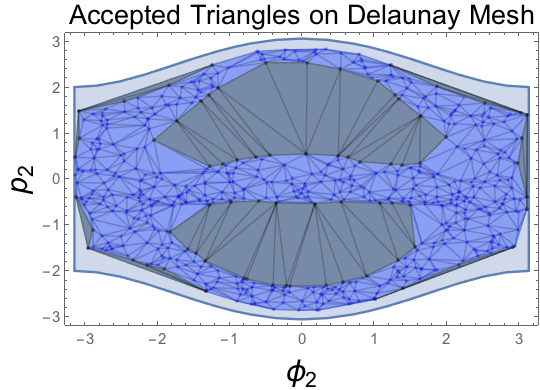}
	\caption{\label{f:accepted-triangles-mesh} Accepted (chaotic, shaded lighter/blue) and rejected (regular, shaded darker/grey) triangles on Delaunay Mesh for a {\it sample} chaotic region on the `$\vf_1=0$' Poincar\'e surface at $E = 7$ for maximal edge length $d=1$. The light colored region on the periphery inside the Hill region consists of regular sections.}
\end{figure}

\begin{figure}
	\includegraphics[width=7.5cm]{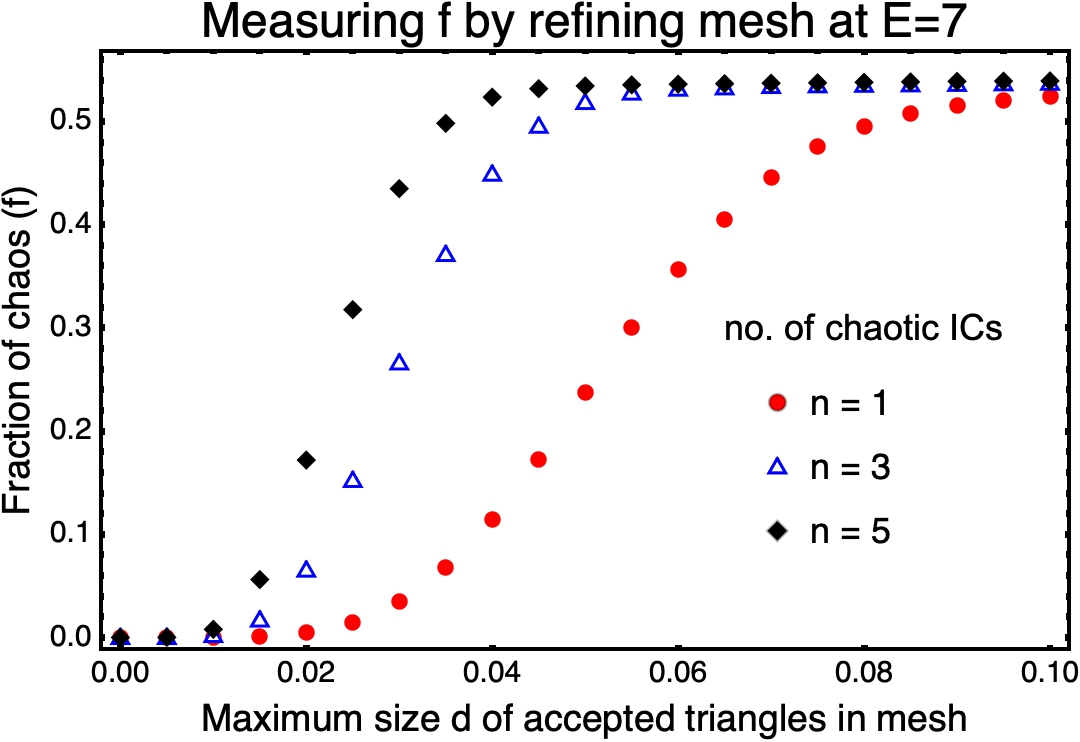}
	\caption{ \label{f:refine-mesh}
Estimates of the fraction of chaos (area of accepted region/area of Hill region) for various choices of  $d$. An optimal estimate for $f$ is obtained by picking $d$ where $f$ saturates. The three data sets displayed have $n=1,3,5$ chaotic ICs, each evolved for the same duration $t = 10^5$.}
\end{figure}

To estimate the fraction of the area of the Hill region (at a given $E$) occupied by the chaotic sections on the `$\vf_1=0$' Poincar\'e surface, we need to assign an area to the corresponding scatter plot (e.g., see Fig \ref{f:psec-egy=4.5}). We use the DelaunayMesh routine in Mathematica to triangulate the scatter plot so that every point in the chaotic region lies at the vertex of one or more triangles (see Fig. \ref{f:accepted-triangles-mesh}). For such a triangulation and a given $d > 0$, the $d$-area of the chaotic region is defined as the sum of the areas of those triangles with maximal edge length $\leq d$ (accepted triangles in Fig. \ref{f:accepted-triangles-mesh}). Fig.~\ref{f:refine-mesh} shows that the area initially grows rapidly with $d$, and then saturates for a range of $d$. Our best estimate for the area of the chaotic region is obtained by picking $d$ in this range. Increasing $d$ beyond this admits triangles that are outside the chaotic region. Increasing the number of points in the scatter plot (either by evolving each IC for a longer time or by including more chaotic ICs, which is computationally more efficient) reduces errors and decreases the threshold value of $d$ as illustrated in Fig. \ref{f:refine-mesh}.


\footnotesize


\end{document}